\documentclass[review,12pt]{elsarticle}

\usepackage{amsmath}
\usepackage{color}
\usepackage{amssymb}
\usepackage{booktabs}
\usepackage{graphicx}
\usepackage{epstopdf}
\usepackage{enumerate}
\usepackage{subfigure}
\usepackage{upgreek}
\usepackage{setspace}
\usepackage{textcomp}

\newcommand{\eqr}[1]{(\ref{eq:#1})}

\newcommand{\pd}[2]{\frac{\partial #1}{\partial #2}}

\newcommand{\eg}{\textit{e.g.}}
\newcommand{\bs}[1]{\boldsymbol{#1}}
\newcommand{\etal}{\textit{et al.\,}}
\newcommand{\ie}{\textit{i.e.}}

\graphicspath{ {./} }

\journal{International Journal of Multiphase Flow}

\begin{document}
\begin{frontmatter}

\title{{Modeling and Detailed Numerical Simulation of the Primary Breakup of a Gasoline Surrogate Jet under Non-Evaporative Operating Conditions}}

\author[add1]{Bo Zhang}
\author[add2]{Stephane Popinet}
\author[add1]{Yue Ling\corref{cor1}}
\ead{Stanley\_Ling@baylor.edu}
\cortext[cor1]{Corresponding author. 
	Address:  
 	One Bear Place \#97356, Waco, TX 76798
 }
\address[add1]{Department of Mechanical Engineering, Baylor University, Waco, Texas 76798, United States}
\address[add2]{Institut Jean Le Rond d'Alembert, Sorbonne Universit\'es, UPMC Univ Paris 06, UMR 7190, F-75005, Paris, France} 

\begin{abstract}
The injection and atomization of gasoline fuels are critical to the performance of gasoline direct injection engines. Due to the complex nature of the primary breakup of the liquid jet in the near field, high-level details are often difficult to measure in experiments. In the present study, detailed numerical simulations are performed to investigate the primary breakup of a gasoline surrogate jet under non-evaporative ``Spray G" operating conditions. {The Spray G injector and operating conditions, developed by the Engine Combustion Network (ECN),} represent the early phase of spray-guided gasoline injection. To focus the computational resources on resolving the primary breakup, simplifications have been made on the injector geometry. The effect of the  internal flow on the primary breakup is modeled by specifying a nonzero injection angle at the inlet. The nonzero injection angle results in an increase of the jet penetration speed and also a deflection of the liquid jet. A parametric study on the injection angle is performed, and the numerical results are compared to the experimental data to identify the injection angle that best represents the Spray G conditions. The nonzero injection angle introduces an azimuthally non-uniform velocity in the liquid jet, which in turn influences the instability development on the jet surfaces and also the deformation and breakup of the jet head.  The asymmetric primary breakup dynamics eventually lead to an azimuthal variation of droplet size distributions. The number of droplets varies significantly with the azimuthal angle, but interestingly, the probability density functions (PDF) of droplet size for different azimuthal angles collapse to a self-similar profile. The self-similar PDF is fitted with both lognormal and gamma distribution functions. Analysis has also been conducted to estimate the percentage and statistics of the tiny droplets that are under resolved in the present simulation. The PDF of the azimuthal angle is also presented, which is also shown to exhibit a self-similar form that varies little over time. The PDF of the azimuthal angle is well represented by a hyperbolic tangent function. Finally, a model is developed to predict the droplet number as a function of droplet diameter, azimuthal angle where a droplet is located, and time.
\end{abstract}

\begin{keyword}
DNS, atomization, gasoline direction injection, droplet size distribution
\end{keyword}

\end{frontmatter}

\section{Introduction} 

A comprehensive understanding of the injection and atomization of gasoline fuels is essential to improving the fuel injection systems in gasoline direct injection (GDI) engines. The characteristics of the droplets formed in the atomization process have a direct impact on the subsequent turbulent dispersion of droplets, droplet evaporation, mixing between the fuel vapor and the air, and eventually combustion features like spark ignition and flame propagation in engines \cite{Zhao_1999a}. Due to the increasing demand for high fuel efficiency and low pollutant emission, extensive research efforts have been directed toward understanding and predicting the atomization of gasoline jets and the resulting spray characteristics in the past decades \cite{Mitroglou_2006a,Wang_2015a,Duke_2017a,Khan_2017a,Sphicas_2018a, Payri_2017a}. {For the purpose of advancing the understanding of gasoline spray formation, the Engine Combustion Network (ECN) has developed the benchmark ``spray G" injector and operating conditions. ECN has also provided a rich experimental database for numerical model validation. In the present study we will develop a numerical model for  a gasoline non-evaporative surrogate jet under the spray G operating conditions and investigate the primary breakup of the liquid jet. }

The breakup or atomization of a liquid jet is usually divided into the primary and secondary breakup/atomization processes: while the former is referred to the disintegration of bulk liquid jets into droplets and ligaments, the latter describes the breakups of large droplets and ligaments to even smaller ones. The primary and secondary breakups can happen simultaneously and the boundary between the two processes is often blurry. The primary breakup typically dominates in the near field and the secondary breakup appears mostly in the mid/far field. The primary breakup  of a liquid jet is a problem of enormous complexity and involves multiple physical processes occurring in a wide range of spatial scales \cite{Reitz_1982a,Lin_1998a,Aleiferis_2010a}. This multi-scale nature makes the investigation of primary breakup challenging. Furthermore, the flow of the liquid fuel inside the injector (\ie, the so-called internal flow) can also affect the breakup dynamics of the liquid jet outside the nozzle \cite{Payri_2016a,Agarwal_2020a}, which further complicates the problem. Experiments have been the major approach to investigate gasoline injection in the past \cite{Mitroglou_2006a, Duke_2017a, Aleiferis_2010a}. However, even with the most advanced optical and X-ray diagnostics, there remain two-phase flow features that are hard to measure in experiments. This is in particular true for the near field where the primary breakup happens  \cite{Heindel_2018a}. As a result, numerical simulation is an important alternative to shed light on the underlying flow physics \cite{Gorokhovski_2008a}. 

Due to the wide range of length scales involved in liquid fuel injection and atomization, a direct numerical simulation (DNS) that can fully resolve all the scales is generally too expensive. The recent rapid development of numerical methods and computer power has enabled large-scale numerical simulations of the primary breakup of a  liquid jet \cite{Fuster_2009a, Lebas_2009a, Desjardins_2010a, Shinjo_2010a, Li_2016a, Ling_2017a, Shao_2017a, Ling_2019a, Hasslberger_2019a}. These simulations adopt the DNS approach, namely solving the Navier-Stokes equations for the interfacial two-phase flows without explicit physical models. Interface-capturing methods, \eg, the volume-of-fluid (VOF) and the level-set methods, were used to resolve the sharp interfaces separating the two immiscible fluids. Ideally, the mesh resolution should be fine enough to fully resolve the turbulence (to the Kolmogorov scale), the interfaces (the surfaces of the smallest droplets) and the interaction between the two. Nevertheless, the minimum cell sizes used in most of these simulations were several microns and thus will not be sufficient to capture the sub-micron droplets that are known to exist from experiments.  The general consensus has been that while the small-scale physics are under-resolved, the large-scale flow remains correct. Since small sub-micron droplets and filaments contain little mass, leaving them under-resolved should have only minor impact on the overall results. Therefore, these ``DNS" simulations should be viewed as high-resolution \emph{detailed numerical simulation} without explicit physical models. There are also studies in the literature which employed sub-grid scale (SGS) model established in single-phase turbulent flows and used  interfacial-capturing methods to resolve the interfaces   \cite{Lakehal_2012a, Agbaglah_2017a}. However, the single-phase SGS models do not account for two important physical processes in atomization: the unresolved morphology or topology changes of the interfaces, and the interaction between turbulence and interfaces. Therefore, the capability of this type of LES approach on capturing the unresolved two-phase turbulence remains to be examined \cite{Aniszewski_2016a}. So far, the best way to examine whether a high-fidelity simulation (HFS), either DNS or LES, truly captures the ``high-fidelity" details is through a grid refinement study, namely examining if the simulation results yield converged or converging results toward high-fidelity experimental data or analytical solutions. For example, the recent DNS study by Ling \etal \cite{Ling_2017a, Ling_2019a} has varied the mesh for four different levels (from 8 million to 4 billion cells) to identify the resolution required to capture converged high-order turbulence statistics (such as turbulent kinetic energy dissipation) in airblast atomization. 

Due to the extreme cost of HFS of atomization, a low-fidelity simulation (LFS) approach is often adopted in macro-scale simulations of practical gasoline fuel injection applications \cite{Dukowicz_1980a, Hoyas_2013a, Aguerre_2019a,Paredi_2020a}. Since the mesh resolution is not enough to resolve the physical process in atomization, including the primary breakup of the liquid jet, micro-scale flows around droplets, secondary breakup, droplet collision and coalescence, and small turbulent eddies, different physical models are then required to represent these unresolved physics. The primary breakup is often modeled in the Lagrangian framework, in which the liquid fuels are injected into the domain as discrete parcels/blobs (one parcel represents multiple physical droplets), instead of a continuous bulk liquid jet \cite{Dukowicz_1980a}. The droplet formation from the primary breakup is considered to be driven by the shear instability, see for example the Kelvin-Helmholtz (KH) instability model; while the droplet secondary breakup is considered to be dictated by the Rayleigh-Taylor accelerative instability. The hybrid KH-RT model for droplet breakup has been widely used in fuel injection simulations, yielding reasonable agreement with experiments \cite{Beale_1999a, Duret_2013a}. Primary breakup models have also been proposed based on the Eulerian framework, such as the Eulerian/Lagrangian Spray Atomization (ELSA) model \cite{Vallet_1999a,Duret_2013a}. Instead of tracing individual parcels, the ELSA model solves an additional transport equation for the surface density. Furthermore, the unresolved turbulent fluctuations and their effects on the mean flow and droplet breakup also need to be considered. Therefore, the primary breakup models (no matter in Lagrangian or Eulerian frameworks) are usually used together with RANS turbulence models \cite{Sparacino_2019a,Duret_2013a}. Since the flow around each individual droplet is not resolved, the drag force and heat transfer models are required to account for the unresolved interaction between the droplets and surrounding gas \cite{Maxey_1983a, Michaelides_1994a, Ling_2016a}, so that the motion and temperature evolution of the droplets can be captured. 

The extreme computational costs still prohibit a DNS for the whole fuel injection process in GDI engines, even with the computer power today. Nevertheless, DNS is still very important to atomization research since they can resolve the interfacial multiphase flows much more accurately and can provide high-level details that are hard to obtain in experiments or LFS. More important, the physical insights and high-fidelity simulation data obtained in DNS can be used to improve the sub-scale models in LFS through physics-based or data-based approaches. The research direction on improving atomization models through DNS results has received increasing attention and good progress has been made in the past decade \cite{Lebas_2009a,Duret_2013a}. 

In the previous studies of DNS of atomization, the inlet conditions for the liquid jet are usually significantly simplified, compared to the liquid fuel jets in GDI engines. For example, the injection velocities used in DNS are usually lower than practical engine conditions and the effect of internal flow on the primary breakup is ignored  \cite{Lebas_2009a, Desjardins_2010a, Shinjo_2010a}. Therefore, even such a simulation can accurately capture the physics of the primary breakup, the process resolved does not faithfully represent the fuel atomization process occurring in GDI engines. The goal of the present study is to accurately model and simulate the primary breakup of a gasoline jet with operating conditions and injector geometry which better represent realistic engine conditions. The Engine Combustion Network (ECN) ``Spray G" benchmark case is thus employed. In particular, we will focus on modeling and simulating the experiment by Duke \etal \cite{Duke_2017a}. 

The ECN spray G injector geometry is configured based on modern gasoline injection systems and the specified operating conditions correspond to non-reacting early phase of spray-guided gasoline injection. The same injector and operating conditions have been used by different experimental groups with different diagnostic techniques \cite{Duke_2017a, Payri_2017a, Piazzullo_2018a, Sphicas_2018a}. The experimental database can be then used to validate numerical model and simulations. Low-fidelity simulations using Lagrangian \cite{Sphicas_2017a, Aguerre_2019a, Di-Ilio_2019a, Paredi_2020a} and Eulerian \cite{Navarro-Martinez_2020a} approaches have been performed to test the breakup models \cite{Aguerre_2019a, Di-Ilio_2019a, Navarro-Martinez_2020a} and to investigate the inter-plume aerodynamics \cite{Sphicas_2017a}.   Recently, attempts have been made to perform LES of primary breakup including the whole injector geometry \cite{Befrui_2016a, Yue_2020a}. Yet due to the high Reynolds and Weber numbers involved, whether the mesh resolutions in these simulations were sufficient to faithfully resolve both the internal flow and the external turbulent sprays remains to be examined. 

In the present study, in order to focus the computational resources on resolving the primary breakup process, the injector geometry will be simplified. Nevertheless, the boundary conditions at the inlet are carefully specified and calibrated based on the X-ray experimental data \cite{Duke_2017a} to capture the dominant effect of the internal flow on the liquid jet breakup. To allow for a direct comparison between the numerical and experimental results, a low-volatility gasoline surrogate is used in the simulation, following the experiment. As a result, evaporation is ignored in the present study. For DNS of primary breakup, it is crucial to resolving the sharp interfaces separating the gas and liquid phases. A geometric volume-of-fluid (VOF) method that conserves both mass and momentum is thus used in the present simulation. The VOF method has been implemented in the open-source multiphase flow solver, \emph{Basilisk}. The details of the numerical methods and the simulation setup will be explained in section \ref{sec:model}. The results will be presented and discussed in section \ref{sec:results} and we will summarize the key findings in section \ref{sec:conclusions}.

\section{Modeling and Simulation Approaches }
\label{sec:model}
\subsection{Governing Equations}
The one-fluid approach is employed to resolve the gas-liquid two-phase flow, 
where the phases corresponding to the liquid and the gas 
are treated as one fluid with material properties that change abruptly across the interface. 
Both the gas and liquid flows are considered as incompressible, so the Navier-Stokes equations 
with surface tension can be written as
\begin{align}
	\rho \left(\pd{u_j}{t} + u_i \pd{u_j}{x_i}\right) & = -\pd{p}{x_j} 
	+ \pd{(2\mu D_{ij})}{x_i} + \sigma \kappa \delta_s n_j\, , 
	\label{eq:mom} \\
	\pd{u_i}{x_i} & = 0 \, ,
	\label{eq:divfree}
\end{align}
where $\rho$, $\mu$, $u$, and $p$ represent density, viscosity, velocity and pressure, respectively, and the subscripts $i,j=1,2,3$ represent the Cartesian indices. 
The deformation tensor is denoted by $D_{ij}=(\partial_i u_j + \partial_j u_i)/2$. 
The third term on the right hand side of Eq.\ \eqr{mom} is a singular term, with 
a Dirac distribution function $\delta_s$ localized on the interface, and it represents 
the surface tension. The surface tension coefficient is $\sigma$, 
and $\kappa$ and $n_i$ are the local curvature and unit normal vector of the interface. The surface tension coefficient $\sigma$ is taken as constant in the present study.

The two different phases are distinguished by a characteristic function $c$, and the temporal evolution of which satisfies the advection equation
\begin{align}
	\pd{ c}{t} + u_i \pd{c}{x_i} =0 \, ,
\end{align}
the conservative form of which can be expressed as
\begin{align}
	\pd{ c}{t}  + \pd{(c u_i)}{x_i} = c \pd{u_i}{x_i} \, ,
	\label{eq:char_func}
\end{align}
For incompressible flow, the term on the right hand side is identical to zero. 

\subsection{Numerical methods}
\label{sec:num_methods}
The momentum-conserving volume-of-fluid (MCVOF) method of Fuster and Popinet \cite{Fuster_2018a} is employed to resolve the interfacial two-phase flows. In the original paper, the method was introduced in the context of compressible flows.  Here we summarize only the important steps that are related to incompressible flows.

\subsubsection{Volume-of-fluid method}
In VOF method, the advection equation for $c$, Eq.\ \eqr{char_func}, is solved in its integral form
\begin{align}
	\Delta \Omega \pd{f}{t}  +  \oint_{\partial \Omega} c u_i n_i \mathrm{d}s =\int_{\Omega} c  \pd{u_i}{x_i}\mathrm{d} V\, ,
	\label{eq:adv_color_func1}
\end{align}
where $\Delta\Omega$ is the cell volume, and $\partial \Omega$ represents the surface of the cell. 
The mean value of $c$ in the cell is denoted by $f$, 
\begin{align}
	f = \frac{1}{\Delta\Omega}\int_{\Omega} c dV\, ,
\end{align}
which represents the volume fraction of liquid in the cell. 
The fluid density and viscosity can then be evaluated as
\begin{align}
	\rho  & = f \rho_l + (1-f) \rho_g \, , 
	\label{eq:density} \\
	\mu  & = f \mu_l + (1-f) \mu_g \, .
	\label{eq:viscosity}
\end{align}
where the subscripts $g$ and $l$ represent the gas and the liquid phases, respectively. 

The discrete form of Eq.\ \eqr{adv_color_func1} on a Cartesian cell can be expressed as
\begin{align}
	\Delta \Omega \frac{f^{n+1}- f^{n}}{\Delta t}  +  \Delta_i F_{f,i} =c_c \pd{u_i}{x_i} \Delta \Omega\, .
	\label{eq:adv_color_func}
\end{align}
The net flux for all three directions is $\Delta_i F_{f,i}=\Delta_1 F_{f,1}+\Delta_2 F_{f,2}+\Delta_3 F_{f,3}$, based on a direction-split advection approach. It has been shown by Weymouth and Yue \cite{Weymouth_2010a} that the term on the right hand side of Eq.\ \eqr{adv_color_func} is important to guarantee exact mass conservation. Furthermore, $c_c$ is the value of $c$ at the cell center, which can be easily evaluated as $c_c=1$ if $f>0.5$ and $c_c=0$ if $f\le 0.5$. The value of $c_c$ must be kept as a constant for all sweep directions. 
The volume-fraction flux $F_{f,i}$ in the direction $i$ is calculated as 
\begin{align}
	F_{f,i} = f_{a} u_{f,i} S \, ,
	\label{eq:vof_flux}
\end{align}
where $u_{f,i}$ is the $i$-component of velocity at the cell surface where the flux is evaluated, and $S$ is the surface area. The fraction of reference fluid that is advected across the cell surface over $\Delta t$ is $f_a$, which  is calculated based on the reconstruction of the interface. Here the piecewise linear interface construction (PLIC) approach is applied \cite{Scardovelli_1999a}. The interface normal is computed by the Mixed-Youngs-Centered (MYC) method \cite{Aulisa_2007a} and the location of the interface in the cell is calculated based on the method of Scardovelli and Zaleski \cite{Scardovelli_2000a}. 

\subsubsection{Momentum advection}
It has been shown in previous studies that, it is important to conserve momentum in the momentum advection near the interface, which is in particular true for cases with large difference between the densities of the two phases \cite{Vaudor_2017a,Fuster_2019a}. The fundamental requirement is to advect the momentum in Eq.\ \eqr{mom} in a manner consistent with the advection of volume fraction in Eq.\ \eqr{char_func}. 

The momentum equation can be rewritten in its conservative form
\begin{align}
	 \pd{\rho u_j}{t} +   \pd{(\rho u_i u_j)}{x_i} & = -\pd{p}{x_j} 
	+ \pd{(2\mu D_{ij})}{x_i} + \sigma \kappa \delta_s n_j\, . 
	\label{eq:mom_cons} 
\end{align}
The discretization of Eq.\ \eqr{mom_cons} is based on the finite-volume approach and the update of velocity from $u_j^n$ to $u_j^{n+1}$ is done in the following steps \cite{Fuster_2018a}
\begin{align}
	 \frac{\big(\rho_l f u_j\big)^{*} - \big(\rho_l f u_j\big)^{n}}{\Delta t} & = - \Delta_i F_{ml,ij} \, , 	\label{eq:mom_adv1}\\ 
	 \frac{\big(\rho_g (1-f) u_j\big)^{*} - \big(\rho_g (1-f) u_j\big)^{n}}{\Delta t} & = - \Delta_i F_{mg,ij} \, , 	\label{eq:mom_adv2}\\ 
	{  u}_j^* & = \frac{\big(\rho_l f u_j\big)^{*}+\big(\rho_g (1-f) u_j\big)^{*}}{\rho_l f^{n+1} + \rho_g (1-f^{n+1})}\label{eq:mom_adv3}\\ 
	 \frac{ u_j^{**} - u_j^*}{\Delta t} & = \frac{1}{\rho}\pd{(2\mu D_{ij})}{x_i}  \, , 	\label{eq:mom_visc}\\ 
	 \frac{ u_j^{***} -  u_j^{**}}{\Delta t} & =\frac{1}{\rho}  \sigma \kappa\pd{f}{x_j}  \, , 	\label{eq:mom_st}\\
	  \frac{ u_j^{n+1} -  u_j^{***}}{\Delta t} & =-\frac{1}{\rho} \pd{p}{x_j}  \, ,  \label{eq:mom_pres}
\end{align}
where Eqs.\ \eqr{mom_adv1}--\eqr{mom_adv3} are for the advection term, and Eqs.\ \eqr{mom_visc}--\eqr{mom_pres} are for the three forcing terms on the right hand side of Eq.\ \eqr{mom_cons} (viscous stress, surface tension, and pressure). The viscous term is discretized by the Crank-Nicholson method. The surface tension term is discretized using a balanced-force approach \cite{Francois_2006a} and the height-function method is utilized to calculate the local interface curvature \cite{Popinet_2009a}. The projection method is used to incorporate the incompressibility condition. The pressure Poisson equation is solved and the pressure obtained is then used in Eq.\ \eqr{mom_pres} to correct the velocity. The numerical methods to compute these three terms (Eqs.\ \eqr{mom_visc}--\eqr{mom_pres}) have been discussed in detail in \cite{Popinet_2009a} and thus are not repeated here. 

In Eqs.\ \eqr{mom_adv1} and \eqr{mom_adv2}, $F_{ml,ij}$ and $F_{mg,ij}$ are the fluxes of the liquid and gas $j$-momentum on cell surfaces normal to the $i$ direction, which is the momentum analogue of $F_{f,i}$ in Eq.\ \eqr{adv_color_func}. To achieve the important feature of momentum conservation, $F_{ml,ij}$ and $F_{mg,ij}$ are calculated to be consistent with the volume-fraction flux  $F_{f,i}$: 
\begin{align}
	F_{ml,ij} & = (\rho_l u_j)_a f_{a} u_{f,i} S \, ,
	\label{eq:liq_mom_flux}\\
	F_{mg,ij} & = (\rho_g u_j)_a (1-f_{a}) u_{f,i} S \, . 
	\label{eq:gas_mom_flux}
\end{align}
where $(\rho_l u_j)_a$ and $(\rho_g u_j)_a$ denote the liquid and gas momentum per unit volume to be advected. Following the method of \cite{Lopez-Herrera_2015a}, $(\rho_l u_j)_a$ and $(\rho_g u_j)_a$ are advected as tracers associated with the volume fraction of the corresponding phase non-diffusively. The Bell-Collela-Glaz (BCG) second-order upwind scheme \cite{Bell_1989a} is used for the reconstruction of $(\rho_l u_j)$ and $(\rho_g u_j)$ in the cell, and the generalized minmod slope limiter is employed to compute the gradient. 

In order to highlight the advantage of the MCVOF method, we have also solved the advection term in the momentum equation using the standard BCG advection scheme \cite{Bell_1989a} as in former studies \cite{Popinet_2009a}. The results obtained by the two different methods will be compared and discussed in {sections \ref{sec:rising}} and \ref{sec:compare_methods}. 

\subsubsection{Numerical solver}
The above numerical methods have been implemented in the open-source adaptive multiphase solver, \emph{Basilisk} \cite{basilisk}. 
In particular, the VOF associated tracer advection method of \cite{Lopez-Herrera_2015a} was implemented in the header file ``vof.h", which is used for momentum advection in ``conserving.h" \cite{basilisk}.
In \emph{Basilisk}, a finite volume approach based on a projection method is used. 
The mass and momentum control volumes are collocated in the spatial discretization, which makes it easier to calculate the momentum flux consistently with the volume-fraction flux.
A staggered-in-time discretization of the volume-fraction/density and pressure leads to a formally second-order accurate time discretization. An octree spatial discretization is used in 3D simulations, which gives a very important flexibility allowing dynamic grid refinement into user-defined regions. The adaptation criterion is based on the wavelet estimate of the discretization error \cite{Hooft_2018a}. 
The parallelization of the solver is done through a tree decomposition approach to guarantee a high parallel performance even if a large number of refinement levels are used. 

\begin{figure}[htbp]
\begin{center}
\includegraphics [width=1.\columnwidth]{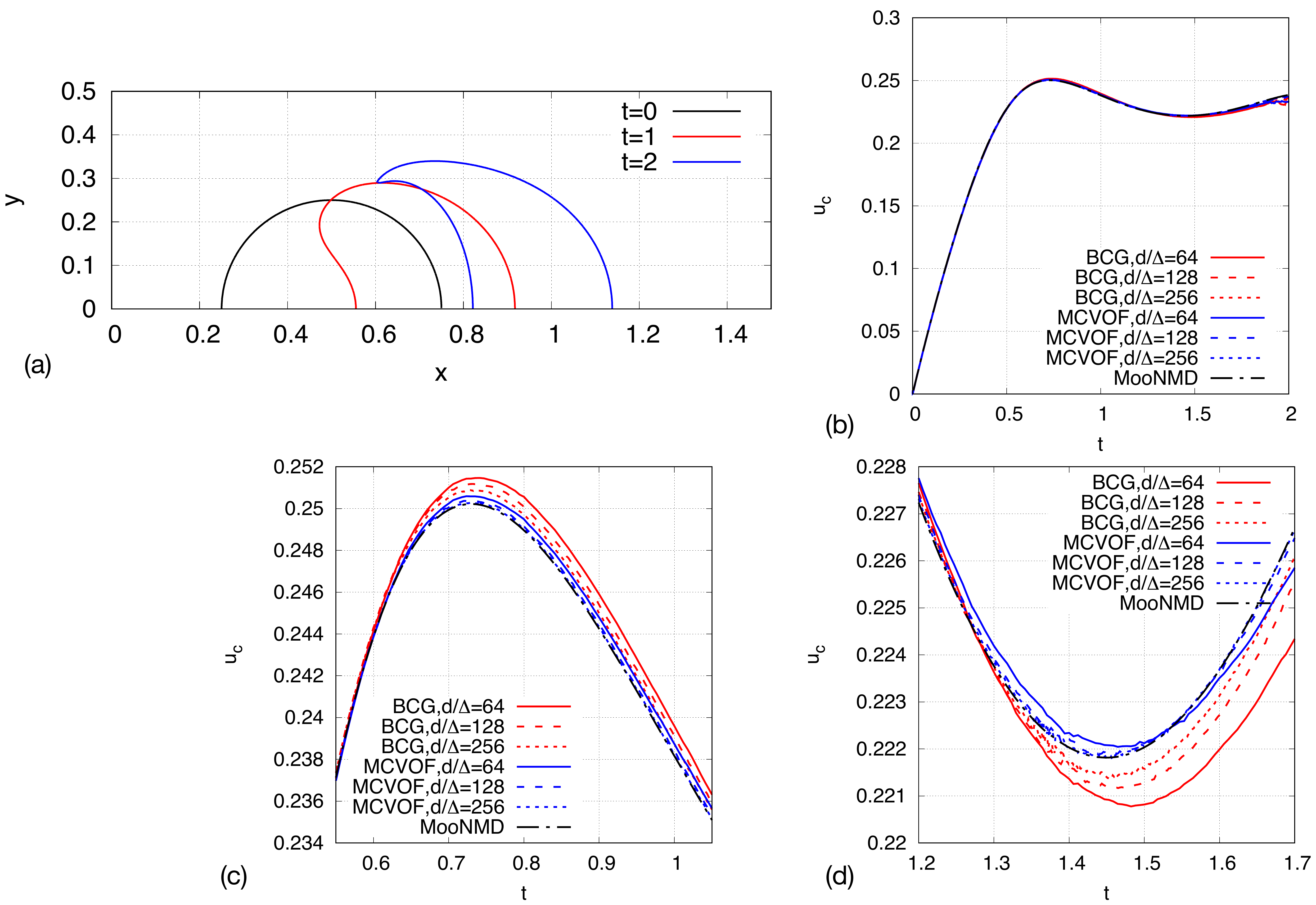}
\end{center}
\caption{Results for the 2D test problem of a rising bubble. (a) Computation domain and bubble surfaces; (b) temporal evolution of bubble centroid velocity; (c) and (d) closeups near the local maximum and minimum of the the bubble velocity.}
\label{fig:rising} 
\end{figure}

{
\subsubsection{Validation test: 2D rising bubble}
\label{sec:rising}
The 2D rising-bubble benchmark problem proposed by Hysing \etal \cite{Hysing_2009a}  is employed to validate the MCVOF method described in section \ref{sec:num_methods} and to examine the distinction between the MCVOF method and the conventional BCG methods. This benchmark case has been tested by different two-phase flow solvers using different numerical methods. The converged numerical results obtained by the MooNMD code \cite{John_2004a,Ganesan_2007a}, which uses an arbitrary Lagrangian--Eulerian approach, can be used as a reference for numerical method validation. The densities and viscosities for the liquid and gas phases are given $\rho_l=1000, \mu_l=10, \rho_g=1, \mu_g=0.1$. The surface tension is $\sigma=1.96$, and the gravity is $g=0.98$. All parameters here are dimensionless. The 2D computational domain and the bubble surfaces at different times are shown in Fig.\ \ref{fig:rising}(a). The bottom of the domain is a symmetric boundary. The bubble is initially a circle of diameter $d=0.25$ and stationary. The bubble rises and deforms due to buoyancy effect. In this test, we have only considered the time up to 2, since capturing the skirt of the bubble formed at later time will require a much higher mesh resolution. The temporal evolution of the bubble centroid obtained by the BCG and MCVOF methods are shown in Fig.\ \ref{fig:rising}(b)-(d). It is observed that the results for both of the methods agree well with the reference data in general.  Closeups at the local maximum and minimum of the centroid velocity are shown in Figs.\ \ref{fig:rising}(c) and (d), which clearly show that the MCVOF method is more accurate and the results converge to the reference data faster when the mesh is refined. It is worth noting that for the coarse mesh ($d/\Delta_{\min}=64$) the MCVOF method does a much better job, compared to the BCG method. This feature is particularly important to atomization simulations, since the mesh resolution is sometimes relatively low in resolving the small-scale interfacial flow features. 
}

\subsection{Modeling and Simulation Setup}
\subsubsection{A simplified model for the spray G injector}
\begin{figure}[tbp]
\begin{center}
\includegraphics [width=\columnwidth]{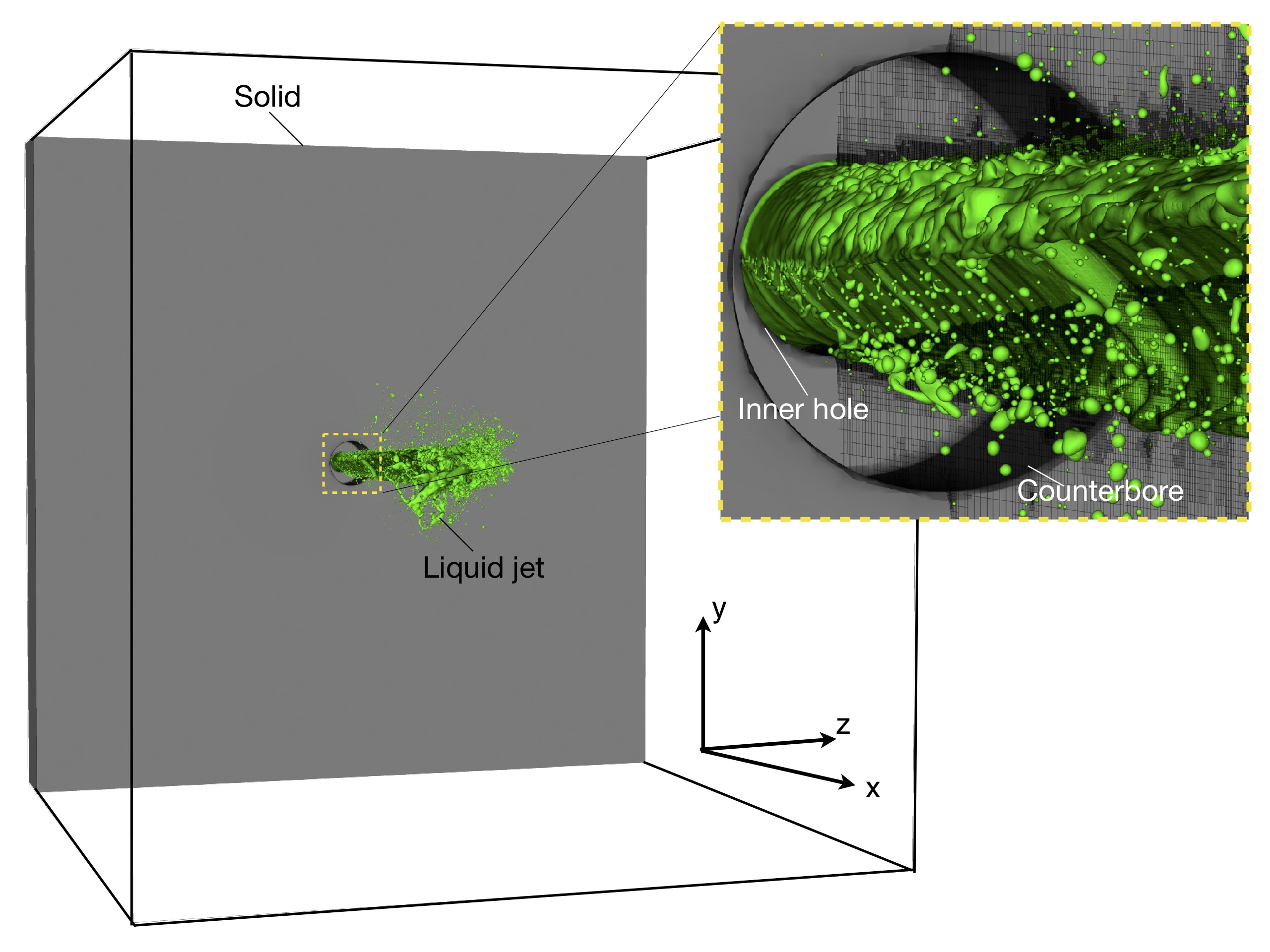}
\end{center}
\caption{Computational domain and the mesh used to simulate the primary breakup of the liquid jet with a nonzero injection angle.}
\label{fig:setup} 
\end{figure}

The computational domain is shown in Fig.~\ref{fig:setup}(a). Simplifications on the injector geometry have been made to focus the computational resources on capturing the interfacial dynamics and primary breakup of the liquid jet. 

First of all, only one of the eight jets generated by the ECN Spray G injector is considered. The original injector has eight holes which are uniformly distributed azimuthally \cite{Duke_2017a}. The jets are spatially separated \cite{Sphicas_2018a}, therefore, ignoring inter-jet interaction will not influence the primary breakup in the near field \cite{Befrui_2016a, Yue_2020a}. 

Furthermore, the injector in the numerical model includes only the inner-hole and counterbore, with the portions upstream, such as the needle, ignored, see Fig.\ \ref{fig:BC}(a). As a result, the internal liquid flow over the needle into the inner-hole will not be simulated. Special boundary conditions, as will be discussed below, will be applied to model the dominant effect of the internal flow on the primary breakup.  

At last, the rate of injection is taken to be a constant. The inlet flow rate in the original Spray G operation varies in time due to the lifting and closing motion of the needle. Here, we only consider the injection rate corresponding to the quasi-steady phase when the needle is completely open. It has been shown in previous experiments that the transition phase is short and its impact on the jet dynamics, such as the penetration length, is generally small \cite{Duke_2017a}.

\begin{figure}[htbp]
\begin{center}
\includegraphics [width=1\columnwidth]{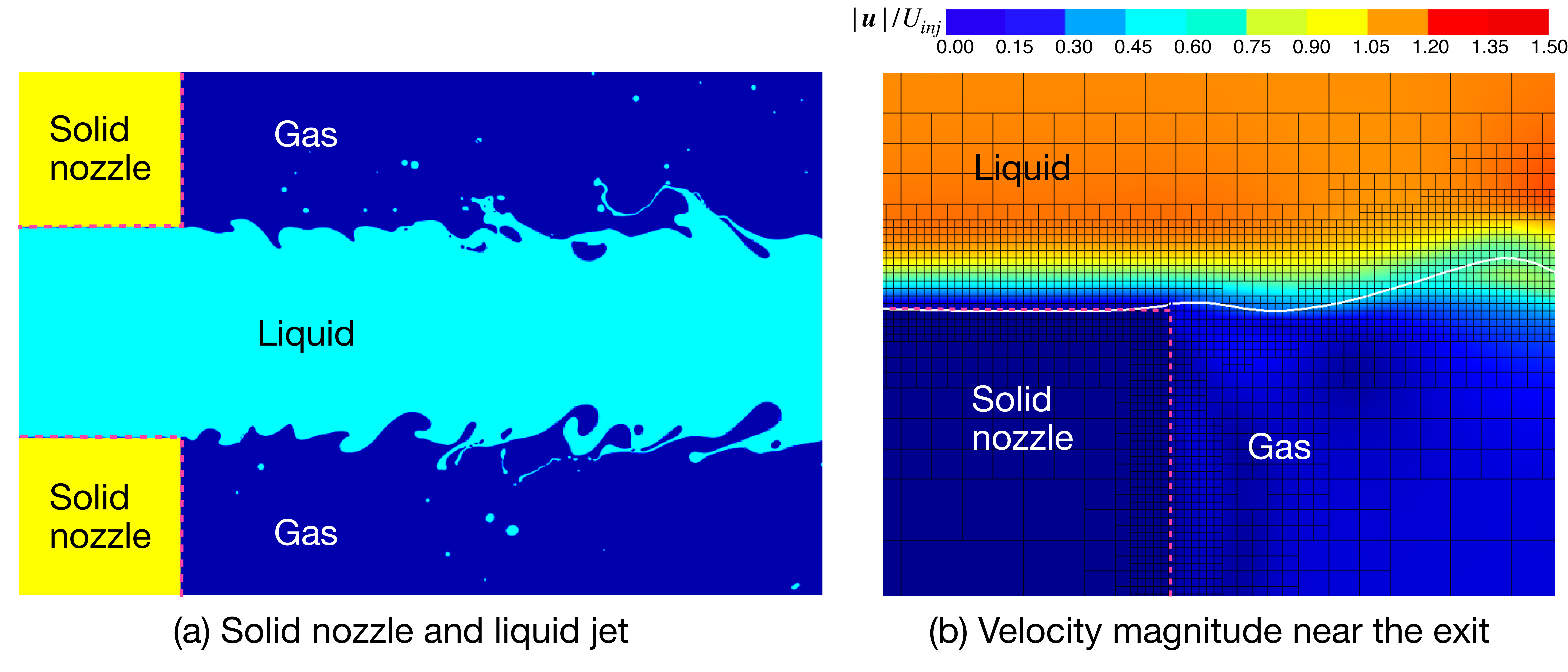}
\end{center}
\caption{(a) The liquid jet at the inner-hole exit and (b) closeup of the velocity field at the nozzle exit. The purple dashed lines indicate the solid boundaries.}
\label{fig:solid} 
\end{figure}

{
The grey color in Fig.~\ref{fig:setup}(a) indicates the embedded solid in the domain, representing the injector geometry. The embedded solid is specified through the solid volume fraction in a cell, $f_s$. Therefore, $f_s=1$ for cells fully occupied by solid, $f_s=0$ for cells with only gas or liquid, and $f_s$ is fractional for cells containing solid boundaries. Since the embedded solid here, namely the injector nozzle, is stationary, the velocity in the cells with $f_s\neq 0$ are masked as $\bs{u} = (1-f_s)\bs{u}$ to achieve the no-slip boundary condition at the solid boundaries. To reduce the numerical error induced by the embedded solid, cells containing solid boundaries are always refined to the maximum refinement level. A 2D test of the liquid jet entering the domain through a solid nozzle was performed and the results are shown in Fig.\ \ref{fig:solid}. As can be seen, the boundary layer near the solid boundary and the gas-liquid interface are well resolved. 
  }

\subsubsection{Boundary Conditions}
\begin{figure}[tbp]
\begin{center}
\includegraphics [width=\columnwidth]{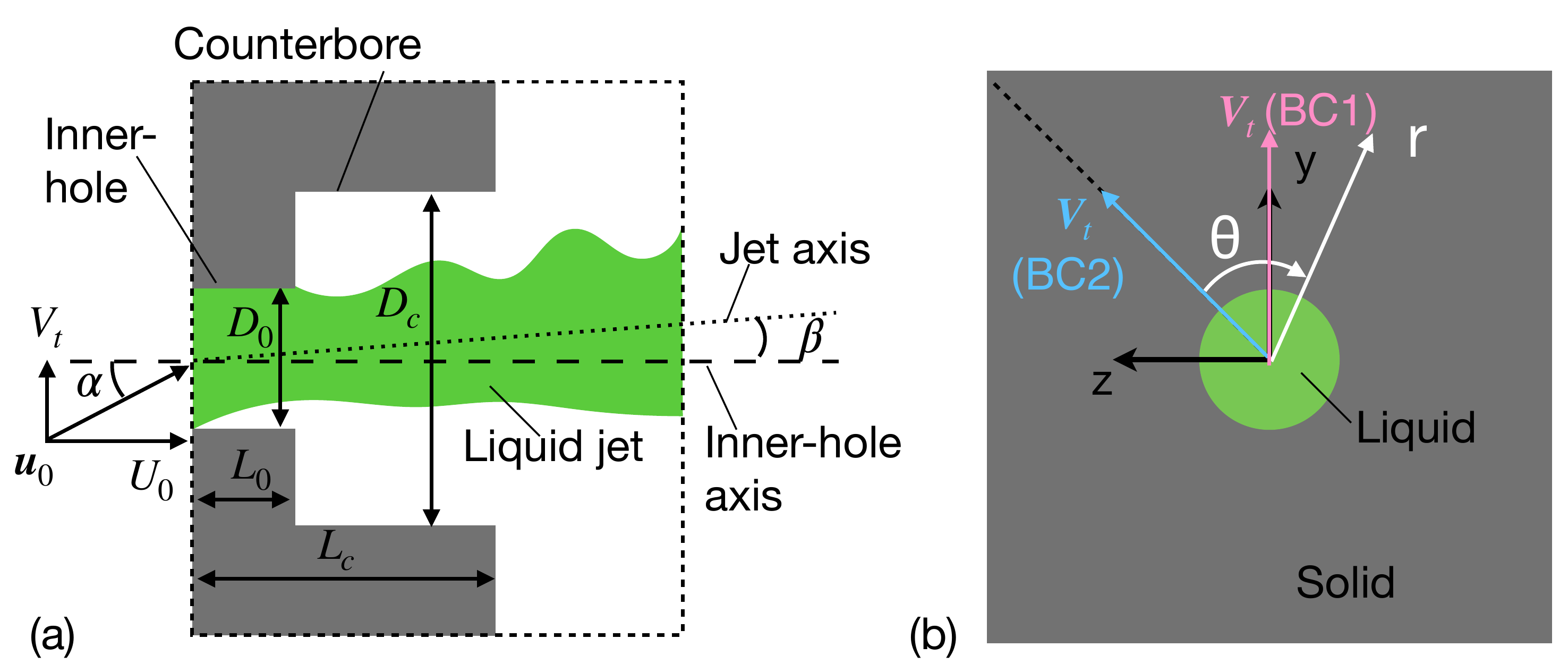}
\end{center}
\caption{Schematics for the inlet boundary conditions on the (a) symmetric plane along the tangential inlet velocity $\bs{V}_t$ and (b) the $y$-$z$ plane at the inlet. Two different ways to specify the tangential inlet velocity are indicated as BC1 and BC2 in (b). }
\label{fig:BC} 
\end{figure}

Previous numerical studies on the full Spray G injector showed that when the liquid flows over the needle and enters the inner-hole,  the liquid velocity at the inlet of the inner-hole is not  aligned with the inner-hole axis \cite{Befrui_2016a}. The angle between the inlet velocity and the inner-hole axis is referred to as the ``injection angle", denoted by $\alpha$. This nonzero injection angle will introduce an interaction between the injected liquid with the inner-hole wall and will influence the macro-scale and micro-scale features of the primary breakup, see the closeup of the jet near the exit of the injector in Fig.~\ref{fig:setup}. In the present study, $\alpha$ is specified through the Dirichlet velocity boundary condition at the inlet, which is schematically shown in Fig.\ \ref{fig:BC}. 

The spatial dimensions of the injector geometry are chosen to be the same as the experiment \cite{Duke_2017a} and are listed in Table \ref{table_argonne}. The normal component of the inlet velocity (along the $x$ axis), $U_0$, is determined by the mass flow rate for the quasi-steady phase of injection \cite{Duke_2017a}. The two tangential components of the inlet velocity, along the $y$ and $z$ axes, are represented by $V_0$ and $W_0$, respectively. The magnitude of the total tangential inlet velocity $V_t=|\bs{V}_t|=\sqrt{V_0^2 + W_0^2}$ varies with the injection angle $\alpha$, or the tangent of $\alpha$, $\eta= \tan (\alpha) =V_t/U_0$. 
We have tested two different ways to specify the tangential inlet velocity $\bs{V}_t$: 1) $V_0=V_t$ and $W_0=0$ and 2) $V_0=V_t/\sqrt{2}$ and $W_0=V_t/\sqrt{2}$. These two boundary conditions are denoted as BC1 and BC2 in Fig.~\ref{fig:BC}(b), respectively. For the BC1, $\bs{V}_t$ is aligned with the y axis and it will be shown later that this exact alignment between $\bs{V}_t$ and the Cartesian mesh will introduce a numerical artifact on the jet surfaces. Rotating $\bs{V}_t$ for 45 degrees as in the BC2 significantly reduces this numerical artifact. 

\begin{table*}[tbp]
\centering
\begin{tabular}{c c c c c c}
    \hline
	$D_0$ & $D_c$ & $L_0$ & $L_c$ & $U_0$ & $V_t$ \\
($\mu$m) & ($\mu$m) & ($\mu$m) & ($\mu$m) & (m/s) & (m/s) \vspace{0.05in}\\
\hline
173 & 388 & 152 & 395 & 89 & 0,\,17.8,\,35.6 \\
    \hline
\end{tabular}
\caption{Dimensions of the inner-hole and counterbore and injection velocity components used in the present simulation. The parameters are chosen to be consistent with the experiment \cite{Duke_2017a}.}
\label{table_argonne}
\end{table*}

For the convenience of discussion of the simulation results, a cylindrical coordinate, $(r,\theta,x)$, is introduced, see Fig.~\ref{fig:BC}(b). The azimuthal angle, $\theta$, is defined with respect to $\bs{V}_t$ according to the BC2. 

{
In the present setup, no disturbance is added in the inlet velocity, yet the numerical error induced by the embedded solid plays the role of inlet flow fluctuations. The turbulent velocity fluctuations at the jet inlet can have an impact in the interfacial instability development and the resulting spray characteristics \cite{Menard_2007a, Jiang_2019c}. A systematic investigation of effect of the inlet disturbance is of interest but out of the scope of the present study. 
}

The pressure-outlet boundary condition is invoked at the right surface of the domain. All lateral boundaries of the domain are taken to be slip walls. Thanks to the adaptive mesh, a large simulation domain is used. The length of the cubic domain edge is $H=32 D_0$, where $D_0$ is the diameter of the inner-hole, see Fig.~\ref{fig:BC}(a). The effects of the lateral boundaries on the jet are negligible. 

\subsubsection{Mesh resolution}
The octree mesh is used to discretize the domain. The local cell size is adapted based on the estimated discretization errors of the volume fraction $f$ and the three components of velocity $u_i$. The assessment of discretization error for each scalar is achieved through a wavelet transform \cite{Hooft_2018a}. If the estimated error is larger than the specified threshold, the mesh will be locally refined, or vice versa. For the present simulation, the normalized error thresholds for the volume fraction and all three velocity components are all set as 0.01. 
{
For the present problem, the mesh is generally refined to the maximum level near the jet surfaces. The error threshold for velocity is used to identify the region away from the jet, where the mesh can be coarsened. As shown in Figs.\ \ref{fig:setup} and \ref{fig:solid}, the threshold values used here are sufficient to refine the mesh to resolve the interfaces and the shear layers near the interfaces. 
}

The minimum cell size in the octree mesh is controlled by the maximum refinement level, $L$, \ie, $\Delta_{\min}=H/2^L$. Two different meshes have been used, $L=11$ ($\Delta_{\min}=2.70$ \textmu m) and $L=12$ ($\Delta_{\min}=1.35$ \textmu m), and the corresponding meshes are denoted as L11 and L12, respectively.  A representative snapshot of the L12 mesh is shown in Fig.~\ref{fig:setup}. It can be seen that a high mesh resolution is used to resolve the jet surfaces and the flow nearby, while the mesh away from the jet is coarsen to reduce the computational cost. The total number of cells increases in time as more and more liquid enters into the domain. The mesh shown in Fig.~\ref{fig:setup} consists of about 160 million cells. The maximum number of cells in the L12 mesh simulation goes up to 210 million, compared to $(2^{12})^3\approx 69$ billion cells for the equivalent uniform Cartesian mesh. The simulations for the L11 mesh were performed on the Baylor cluster \emph{Kodiak} using 144 cores (Intel E5-2695 V4). The simulation for the L12 mesh was run on the machine \emph{Stampede2} at the Texas Advanced Computing Center with 1440 cores (Intel Xeon Platinum 8160) for about 4 days. 

\subsection{Fluid properties and key parameters}
The fluids properties and the injection conditions are chosen to be similar to the experiment by Duke \etal \cite{Duke_2017a}. The X-ray diagnostics facilities at Argonne National Laboratory were used in the experiment and were restricted to non-evaporative conditions. Therefore, the liquid and gas were replaced by a low-volatility gasoline surrogate (Viscor 16br, Rock Valley Oil \& Chemical Company) and nitrogen, respectively. The chamber pressure was decreased so that the gas-to-liquid density ratio remains the same as the standard Spray G conditions. 

\begin{table*}[tbp]
\centering
\begin{tabular}{c c c c c}
    \hline
$\rho_l$ & $\rho_g$ & $\mu_l$ & $\mu_g$ & $\sigma$ \\
(kg/m$^3$) & (kg/m$^3$) & (Pa s) & (Pa s) & (N/m) \vspace{0.05in}\\
\hline
838 & 3.6 & $9.64 \times 10^{-4}$ & $1.77 \times 10^{-5}$ & 0.0278 \\
    \hline
\end{tabular}
\caption{Fluid properties used in the simulation. The parameters are chosen to be consistent with the experiment by Duke \etal \cite{Duke_2017a}.}
\label{table_argonne}
\end{table*}

\begin{table*}[tbp]
\centering
\begin{tabular}{c c c c c}
    \hline
$Re_g$ & $Re_l$ & $We_l$ & $\xi$ & $\eta$  \vspace{0.05in}\\
$D_0U_0/\nu_g$ & $D_0U_0/\nu_l$  & $\rho_l D_0U_0^2/\sigma$ & $\rho_l/\rho_g$ & $V_0/U_0$\vspace{0.05in}\\
\hline
$3130$ & $13400$ & $41300$ & $233$ & 0, 0.2, 0.4\\
    \hline
\end{tabular}
\caption{Key dimensionless parameters.}
\label{table_parameter}
\end{table*}

If the gas density $\rho_g$, the inner-hole diameter $D_0$, and the normal inlet velocity $U_0$ are chosen to be the reference scales, the key dimensionless parameters can be defined and the values are given in Table~\ref{table_parameter}. The Reynolds and Weber numbers of the liquid jet are defined as $\mathrm{Re}_l=\rho_l (D_0)U_0/\mu_l$ and $\mathrm{We}_l=\rho_l (D_0)U_0^2/\sigma$. For the large values of $\mathrm{Re}_l$ and $\mathrm{We}_l$ here, the viscous and surface tension forces are insufficient to hold the injected liquid as a bulk, and the liquid jet will break. The Reynolds number based on gas properties, $\mathrm{Re}_g=\rho_g D_0U_0/\mu_g$, is defined to characterize the  gas flow induced by the liquid jet. When $\mathrm{Re}_g$ is large, the gas flow will turn to turbulent. The liquid-to-gas density ratio is represented by $\xi$ with $\xi=\rho_l/\rho_g$. Finally, the angle between the inlet velocity and the inner-hole axis is characterized by its tangent, $\eta=\tan\alpha$, and different values of $\eta$ are considered. 

\subsection{Summary of simulation cases}
To investigate the effects of simulation approaches on the results, four different tests have been performed, which are summarized in Table ~\ref{tab:cases}. 
Tests 1 to 3 are done on the coarser L11 mesh to examine the effects of inlet boundary condition (BC1 and BC2) and the numerical method for momentum advection (MCVOF and BCG) on the simulation results. Test 4 uses the same numerical method and boundary condition as Test 3, but is performed on the finer L12 mesh, to show the effect of mesh resolution. For Test 3, different $\eta$, varying from 0 to 0.4 are simulated. The simulation results for these tests will be presented and discussed in section \ref{sec:results}. 

\begin{table*}[tbp]
\centering
\begin{tabular}{c c c c}
    \hline
Test & Maximum level & Boundary Conditions &	Momentum advection method\\
\hline
$1$ & 11 & BC1 & MCVOF\\
$2$ & 11 & BC2	& BCG\\
$3$ & 11 & BC2	 & MCVOF\\
$4$ & 12 & BC2	 & MCVOF\\
    \hline
\end{tabular}
\caption{Test cases for different mesh resolutions, boundary conditions, and momentum-advection methods, considered in the present study.}
\label{tab:cases}
\end{table*}

\section{Results}
\label{sec:results}

\subsection{General effect of the nonzero injection angle on the liquid jet}

\begin{figure}[tbp]
\begin{center}
\includegraphics [width=0.99\columnwidth]{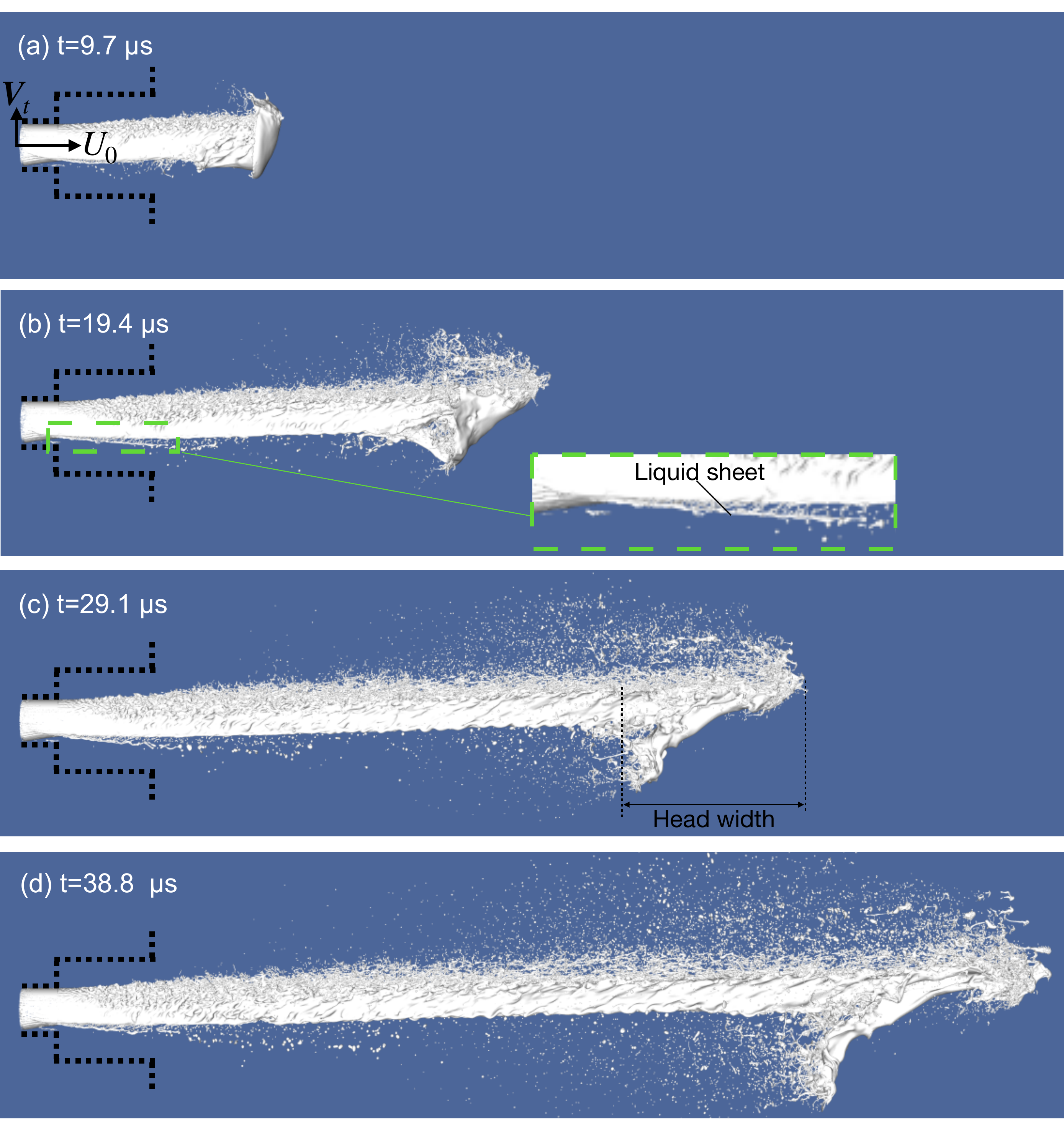}
\end{center}
\caption{Temporal evolution of the liquid jet for $\eta=0.2$ and Test 4 (L12 mesh). The dashed lines denote the boundaries of the inner-hole and counterbore on the central plane.}
\label{vof_evolution} 
\end{figure}

The simulation results for Test 4 and $\eta=0.2$ are shown in Fig.~\ref{vof_evolution} to illustrate the effect of the nonzero injection angle on the liquid jet. In Fig.~\ref{vof_evolution}, the liquid is injected into the stagnant gas from the left, with a view angle for which $\bs{V}_t$ points upward. The boundaries of the inner-hole and counterbore on the central plane are indicated by the black dashed lines. The nonzero injection angle induces several new features of primary breakup that have not been observed in a round jet with zero injection angle \cite{Lebas_2009a, Shinjo_2010a}. 

First of all, the liquid jet is seen to detach from the bottom wall of the inner-hole. The ambient gas is then entrained into the gap between the liquid surface and the inner-hole wall. This phenomenon has also been observed in simulations for the full spray G injector \cite{Befrui_2016a}. 

Secondly, the liquid jet loses its azimuthal symmetry. For the case with zero injection angle, see \eg, \cite{Shinjo_2010a}, the overall shape of the jet remains symmetric, though small-scale features, like interfacial waves and ligaments, may vary azimuthally. Here, the interfacial instability develops much faster on the top surface of the jet than the lateral and bottom surfaces. Furthermore, the top of the jet head moves faster than the bottom, resulting in a stretching of the jet head in the streamwise direction, see Fig.\ \ref{vof_evolution}(c). The upper part of the jet head also breaks earlier and more violently. The asymmetry breakup dynamics eventually leads to a non-uniform spatial distribution of droplets: significantly more droplets are formed above the jet than below. 

At last, it is observed that liquid sheets develop on the two lateral sides of the liquid jet after it leaves the inner-hole, see the closeup in Fig.\ \ref{vof_evolution}(b). This is due to the interaction between the liquid flow and  the inner-hole wall and the resulting flow around the inter-hole wall from the top to the bottom (both clockwise from $\theta=0$ to $\pi$ and also counter-clockwise from $\theta=0$ to $-\pi$). Capillary breakups occur near the edge of this liquid sheet, forming relatively large droplets below the jet. 

\begin{figure}[tbp]
\begin{center}
\includegraphics [width=1\columnwidth]{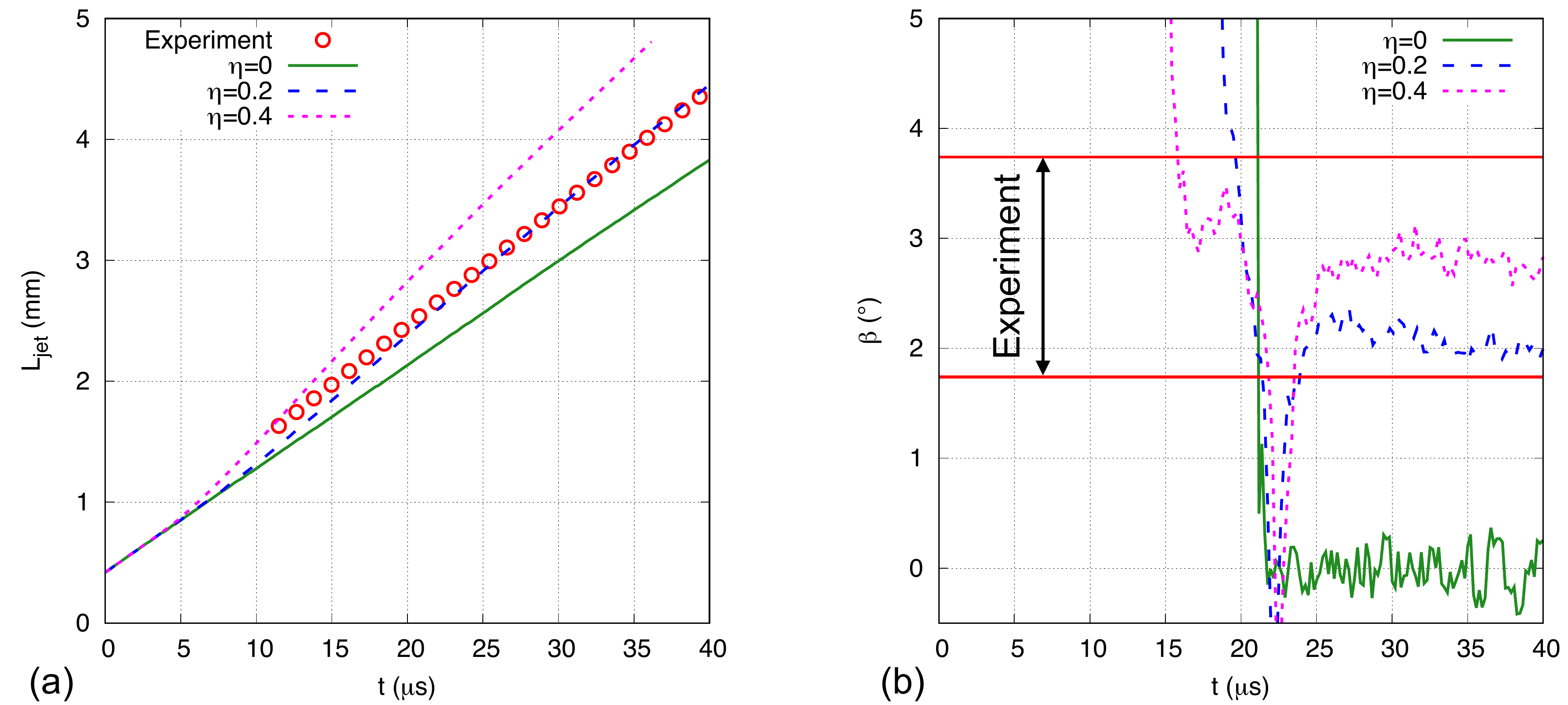}
\end{center}
\caption{Temporal evolutions of (a) the liquid jet penetration and (b) the jet deflection angle for different injection angles, $\eta=\tan(\alpha)$. The simulation results are for Test 3 and the experimental data are from Ref.~\cite{Duke_2017a}.}
\label{fig:penetration} 
\end{figure}

\subsection{Effect of the injection angle on jet penetration and deflection}
\label{sec:angle}
The detachment of the liquid jet within the inner-hole reduces the cross-section area of the liquid jet. Due to mass conservation, the liquid velocity increases, resulting in a faster penetration of the liquid jet.  A quantitative evaluation of the effect of $\eta$ on the jet penetration length is shown in Fig.~\ref{fig:penetration}(a). In order to directly compare the simulation results with the experimental data, the penetration length of the liquid jet, $L_{jet}$, is defined based on the transverse integrated mass (TIM) \cite{Duke_2017a}. The TIM is calculated by integrating the liquid density over the $y$-$z$ plane at a given streamwise location and thus is a function of $x$ and $t$: 
\begin{equation}
     \text{TIM}(x,t) = \iint \rho_l(x,y,z,t) \,\mathrm{d}y\,\mathrm{d}z\, .
     \label{eq:TIM_func}
\end{equation} 
The threshold of TIM for determining the penetration length is taken to be 20\% of TIM$_\mathrm{inlet}$, consistent with the experiment. Results for three different injection angles are shown here, $\eta=0, 0.2$ and 0.4. The slopes of the lines represent the penetration speed. It can be observed that for $\eta=0$, penetration speed is constant. The penetration speed for $t\lesssim 5$ \textmu s varies little with $\eta$ due to the confinement effect of the inner-hole wall. Yet soon after the jet head leaves the inner-hole exit, the penetration speed for nonzero $\eta$ transits to a larger value at about $t =5$ \textmu s. Since then the penetration speed remains unchanged in the rest of the time range considered ($5\lesssim t \lesssim 40$ \textmu s). In the long term, the penetration speed of the jet will decrease in the far field \cite{Duke_2017a}. Nevertheless, the present simulation focuses on the short-term dynamics of the jet in the near field, the variation of the penetration speed the early transition is negligibly small. For convenience, hereafter, we simply refer to the penetration speed as the value after the transition. It can be observed that the penetration speed monotonically increases with $\eta$. The penetration length for $\eta=0.2$ agrees well with the experimental results. 

The nonzero injection angle also induces a deflection of the liquid jet. The deflection angle $\beta$ is defined as the angle between the axes of the liquid jet and the inner-hole, see Fig.\ \ref{fig:BC}(a). The axis of the liquid jet consists of centroids of the liquid phase on the cross sections normal to the $x$-direction. The deflection angle is then calculated as $\beta = \tan^{-1}(\sqrt{y_m^2+z_m^2}/x_m)$, where $x_m,y_m$ and $z_m$  are the coordinates of the centroid of liquid phase. We measured $\beta$ at about $x/D_0=11$ (x=2 mm), following the experiment \cite{Duke_2017a}, and the results are shown in Fig.\ \ref{fig:penetration}(b). The deflection angle can only be measured after the jet has reached the measurement location. The fluctuations for $t=16$ to $24$ \textmu s in the results  are due to the passage of the jet head. After the transition, $\beta$ reaches a quasi-steady state with small-amplitude fluctuations due to the interfacial waves on the jet surface. For $\eta=0$, the mean of $\beta$ is close to zero, namely there is no deflection of the liquid jet. Similar to the penetration speed, the mean of $\beta$ increases monotonically with $\eta$. The experimental result for $\beta$ has a quite large error bar, which is indicated by the two horizontal lines in Fig.\ \ref{fig:penetration}(b). The simulation results for both $\eta=0.2$ and 0.4 lie in the range of the experimental data \cite{Duke_2017a}. The deflection angle $\beta$ is generally smaller than the injection angle $\alpha$ due to the constraint of the inner-hole wall. 

Since the injection angle $\alpha$ is used here to model the dominant effect of the neglected internal flow on the dynamics of the liquid jet, the value of $\alpha$ is not known a priori. The results presented in Fig.~\ref{fig:penetration} serve to identify the value of $\alpha$ that best represents the overall dynamics of the Spray G jet. It is shown that $\eta=0.2$ ($\alpha=11.3$\textdegree) yields the best agreement with the experimental results for both the jet penetration and deflection. More different values of $\eta$ have been tested to identify the best $\eta$ value, though only three of them are shown here. 

\subsection{Effects of simulation approaches on resolving the primary breakup } 
\label{sec:compare_methods}
\begin{figure}[tbp]
\begin{center}
\includegraphics [width=1.\columnwidth]{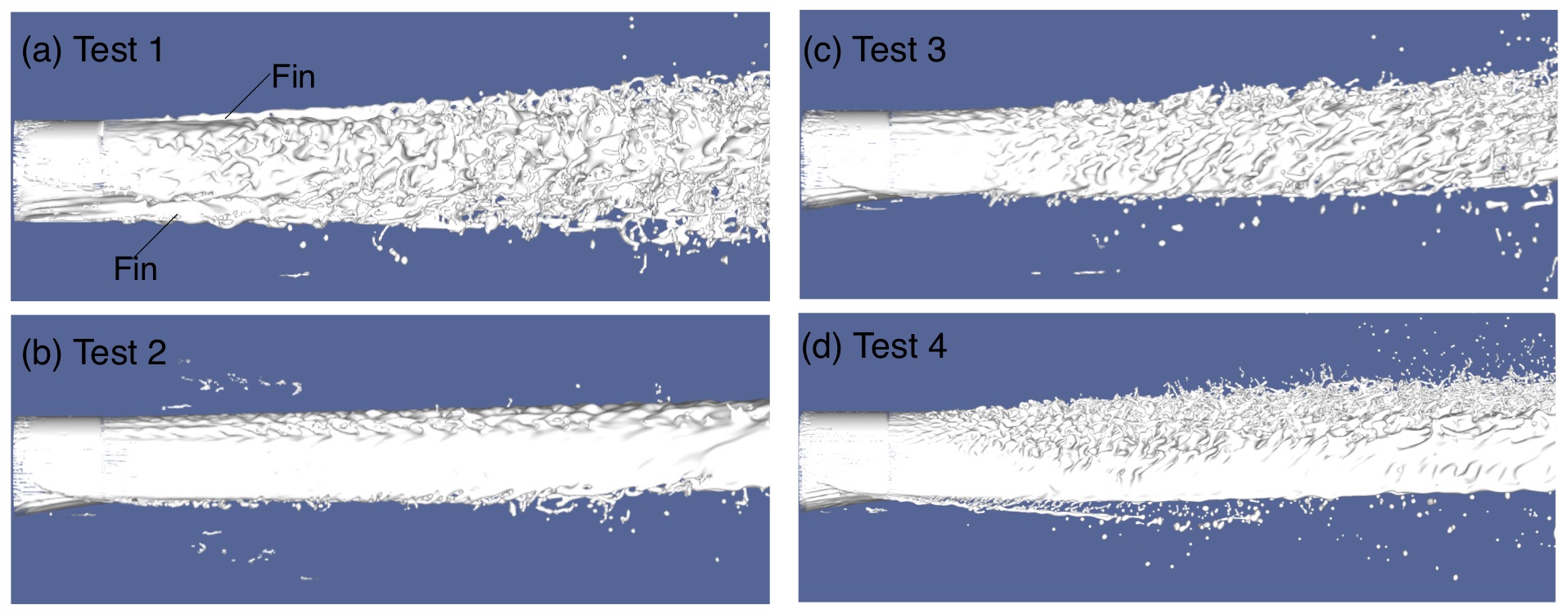}
\end{center}
\caption{The surfaces of the liquid jet at $t=19.4$ \textmu s for different test cases (see Table \ref{tab:cases}) for $\eta=0.2$. (a) Test 1: using the L11 mesh, the boundary condition with the tangential inlet velocity aligned with the $y$-axis (BC1), and the MCVOF method for momentum advection; (b) Test 2: using the L11 mesh, the boundary condition with the tangential inlet velocity rotated 45 \textdegree (BC2), and the BCG method for momentum advection; (c) Test 3: using the L11 mesh, the BC1, and the MCVOF method for momentum advection; (d) Test 4: using the L12 mesh, the BC1, and the MCVOF method for momentum advection.}
\label{compare_vof} 
\end{figure}

\begin{figure}[tbp] 
\begin{center}
\includegraphics [width=0.99\columnwidth]{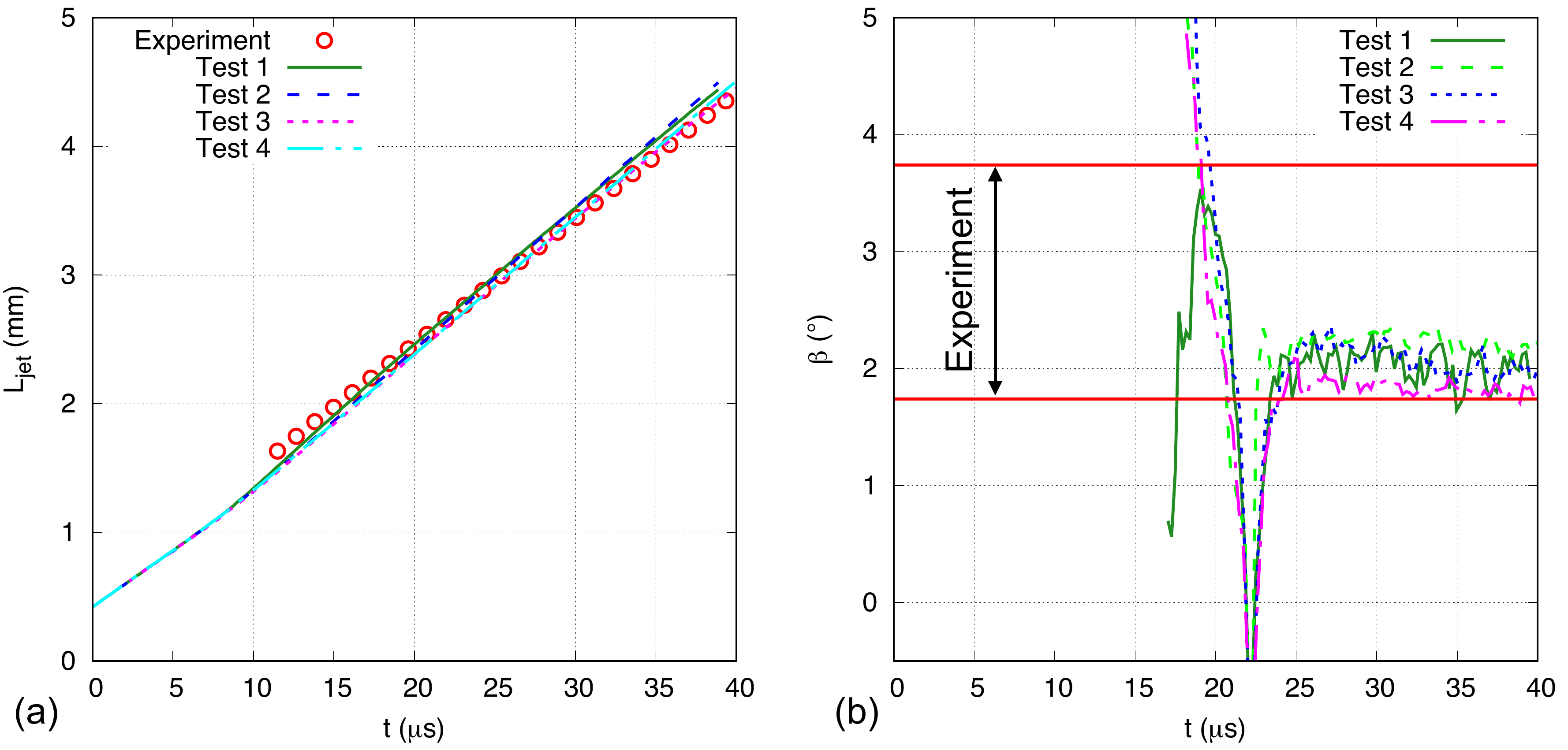}
\end{center}
\caption{Temporal evolutions of (a) the liquid jet penetration and (b) the jet deflection angle for different test cases (see Table \ref{tab:cases}) for $\eta=0.2$. The experimental data are from Ref.~\cite{Duke_2017a}.}
\label{fig:penetration2} 
\end{figure}

To show {that} the simulation approach taken in the present study, in terms of boundary conditions, numerical methods, and mesh resolution, is able and necessary to resolve the primary breakup of the liquid jet with a nonzero injection angle, four different test cases have been performed for $\eta=0.2$, see Table \ref{tab:cases}. The results for the four test cases are shown in Figs.\ \ref{compare_vof} and \ref{fig:penetration2}.

Two different boundary conditions (BC1 and BC2) for the tangential inlet velocity, $\bs{V}_t$, were used in Tests 1 and 3 (see Fig.\ \ref{fig:BC} and Table \ref{tab:cases}). Comparing Fig.\ \ref{compare_vof}(a) and (c), it can be observed that ``fins" are formed on the top and bottom of the jet for Test 1, which is obviously a numerical artifact. Since a Cartesian mesh is used to resolve a cylindrical jet, the numerical error adherent to the Cartesian grid (such as that in the curvature and surface tension calculations) will influence the interfacial instability development. For Test 1, the numerical error is {amplified} due to the alignment of $\bs{V}_t$ with the mesh. In Test 3, the tangential inlet velocity is rotated for 45 degrees, significant improvement was observed and the numerical ``fin" vanishes, see Fig.\ \ref{compare_vof}(c). 

The MCVOF method describe in \ref{sec:num_methods} has been used for momentum advection in the present simulations. {As already shown in section \ref{sec:rising}, the MCVOF method performs better than the BCG method, in particular when the mesh is relatively coarse.} To further evaluate the effect of the momentum-advection method on the primary breakup dynamics, a simulation using purely the BCG method for momentum advection (Test 2) is conducted and the results are compared to those obtained by the MCVOF method (Test 3). The same VOF method has been used to advect the liquid volume fraction for both cases, so the differences in the results are purely induced by the different methods for the momentum advection. 
It can be clearly seen in Figs.\ \ref{compare_vof}(b) and (c) that the jet surfaces for Tests 2 and 3 are very different.  In Test 3, the interfacial waves, the rims and fingers formed at the edges of liquid lobes are captured; while these important primary breakup features are missed in Test 2. Former studies have shown that, a non-momentum-conserving VOF method could introduce numerical breakups of the interfacial waves,  which occur earlier and in smaller spatial scale than the physical reality \cite{Ling_2017a}. The results for Test 2 shown in Fig.\ \ref{compare_vof}(b) correspond to the jet surface after those numerical breakups occurred and that is why the surfaces appear to be smoother than Test 3. 
{
Comparing the results for Tests 2 and 3 (L11 mesh) with those for Test 4 (L12 mesh), it is obvious that the MCVOF results (Test 3) are closer to the fine mesh results. The differences in the results for the jet surface deformation and breakup, captured by the two different numerical methods, will also impact the resulting droplet statistics. 
}

The results for Tests 3 and 4 show the effect of mesh resolution on the primary breakup features. As shown in Fig.\ \ref{compare_vof}(d), Test 4 has captured the smaller wavy structures and ligaments that are not resolved in Test 3. As a result, the formation of smaller droplets is better captured and significantly more droplets are observed in Test 4 than in Test 3. The formation and subsequent breakup of the liquid sheets on the lateral sides of the jet near the inner-hole exit are clearly seen in Test 4, but not in Test 3. This indicates that a fine mesh is necessary to resolve the fine details of the primary breakup and to achieve accurate droplet statistics. Based on the difference between the Tests 3 and 4 results, a simulation with an additional level of grid refinement, \ie, L13, may be needed to fully confirm mesh independency of the simulation results. Due to the high computational cost required, such a simulation will be relegated to our future work. 

It is worth indicating that, the penetration length and the jet deflection angle for these four tests are actually very similar, see Fig.\ \ref{fig:penetration2}. When the mesh is refined from L11 to L12, the jet penetration length and deflection angle vary little, see Fig.\ \ref{fig:penetration2}, and both agree well with the experimental results. Similar conclusions can be made for the change of boundary conditions and numerical methods. This observation seems to show that the micro-scale breakup features do not have a strong influence on the macro-scale dynamics of the jet. Nevertheless, a high mesh resolution, proper boundary condition setup, and accurate numerical methods are required to resolve the micro-scale features like interfacial waves and formation of ligaments and droplets. 

The results in sections \ref{sec:angle} and \ref{sec:compare_methods} have affirmed that, the numerical model for the injection angle $\eta=0.2$ and the simulation approaches specified in Test 4 will capture both the macro-scale and micro-scale primary breakup features of the liquid jet. Therefore, in the rest of the paper, we will focus on the results for $\eta=0.2$ and Test 4. 

\subsection{Interfacial waves on the jet core}
\begin{figure}[tbp]
\begin{center}
\includegraphics [width=.99\columnwidth]{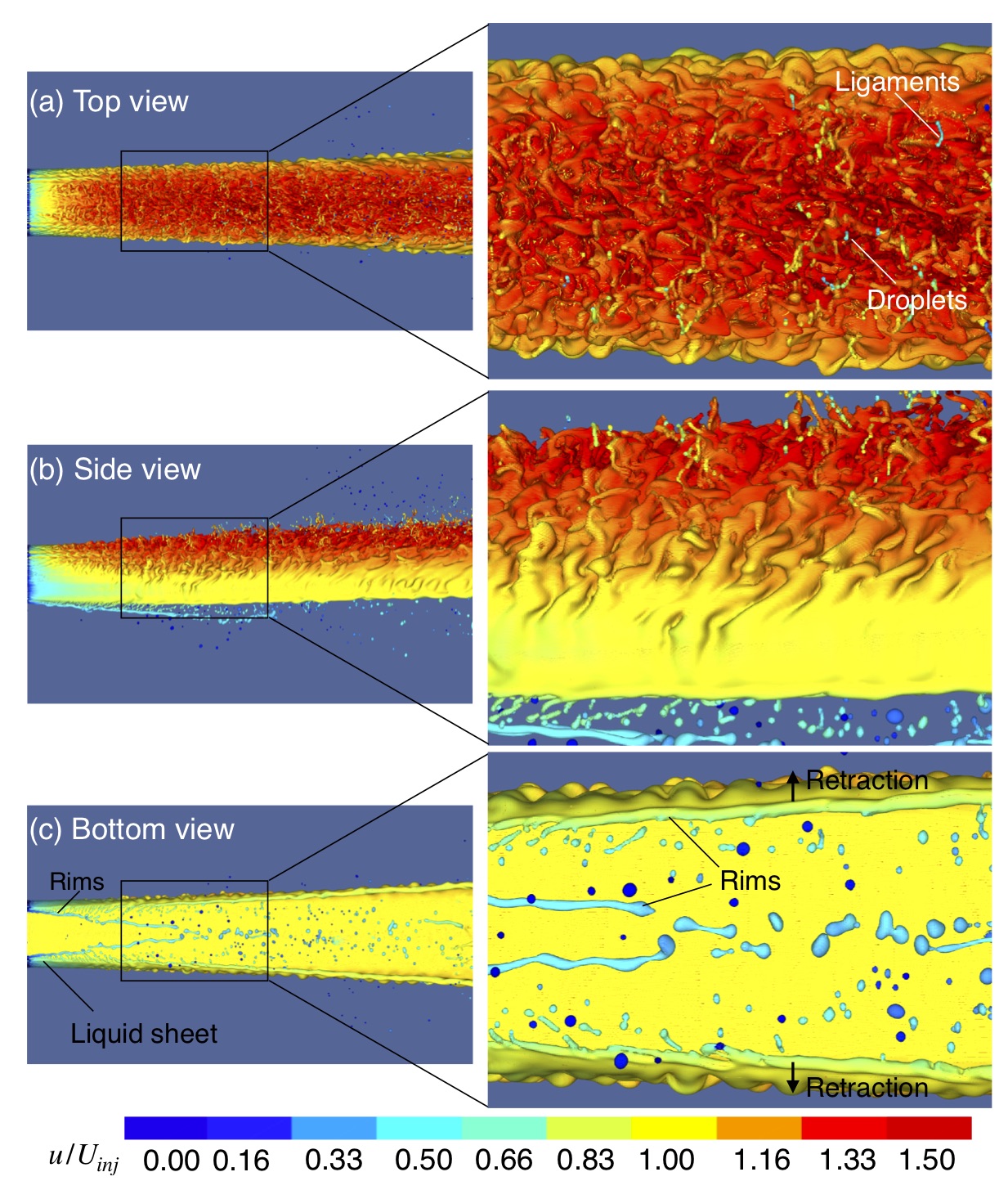}
\end{center}
\caption{Interfacial waves on the jet core surface at $t=19.4$ \textmu s.  The gas-liquid interfaces are colored by the streamwise velocity $u$.  }
\label{velocity1} 
\end{figure}

The liquid jet surfaces at $t=19.4$ \textmu s near the inner-hole exit are shown  in Fig.\ \ref{velocity1} from different view angles. The gas-liquid interfaces are colored with the streamwise velocity. At this time, the portion of the jet shown ($x/D_0\lesssim 7 $) has reached a statistically steady state, namely the average features of the surface morphology and the streamwise velocity do not vary in time. 

The color on the jet surface clearly shows that the streamwise velocity is higher at the top of the jet ($\theta=0$) and decreases clock-wisely from $\theta=0$ to $\pi$ (also counter-clock-wisely from $\theta=0$ to $-\pi$ due to symmetry). Since the shear interfacial instability on the jet surface is driven by the velocity difference between the liquid and gas \cite{Squire_1953a, Yih_1967a,Otto_2013a}, the larger velocity at the top of the jet results in faster growing longitudinal interfacial waves. As the waves are advected downstream and grow in amplitude, the transverse waves arise and develop into lobes or fingers \cite{Marmottant_2004a, Jarrahbashi_2016a}. Following the longitudinal waves, the transverse waves and lobes/fingers also develop faster at the top of the jet. The lobes/fingers are stretched by the surrounding gas and eventually disintegrate into small ligaments and droplets. After the ligaments and droplets are detached from the jet core, the aerodynamic drag causes them to slow down, as indicated by the blue color of the droplets and ligaments above the jet shown in the closeup of Fig.\ \ref{velocity1}(a). 

Due to the nonzero injection angle and the interaction between the injected liquid and the inner-hole wall, liquid sheets are formed on the two lateral sides of the jet near the inner-hole exit and extend toward the bottom, see Figs.\ \ref{velocity1}(b) and (c). Holes arise in the liquid sheet soon after the liquid exits the inner-hole, which cause the liquid sheet to rapture. The rims at the edges of the sheets are then separated from the jet core and become long ligaments. The unbroken liquid sheets attached to the jet core retract back toward the jet due to the Taylor-Culick effect.  The two rims detached from the jet core, at the center of Fig.\ \ref{velocity1}(c), eventually break into droplets. These droplets are significantly larger than those formed from the interfacial waves at the top of the jet, see Fig.\ \ref{velocity1}(b). 

\begin{figure}[tbp]
\begin{center}
\includegraphics [width=.99\columnwidth]{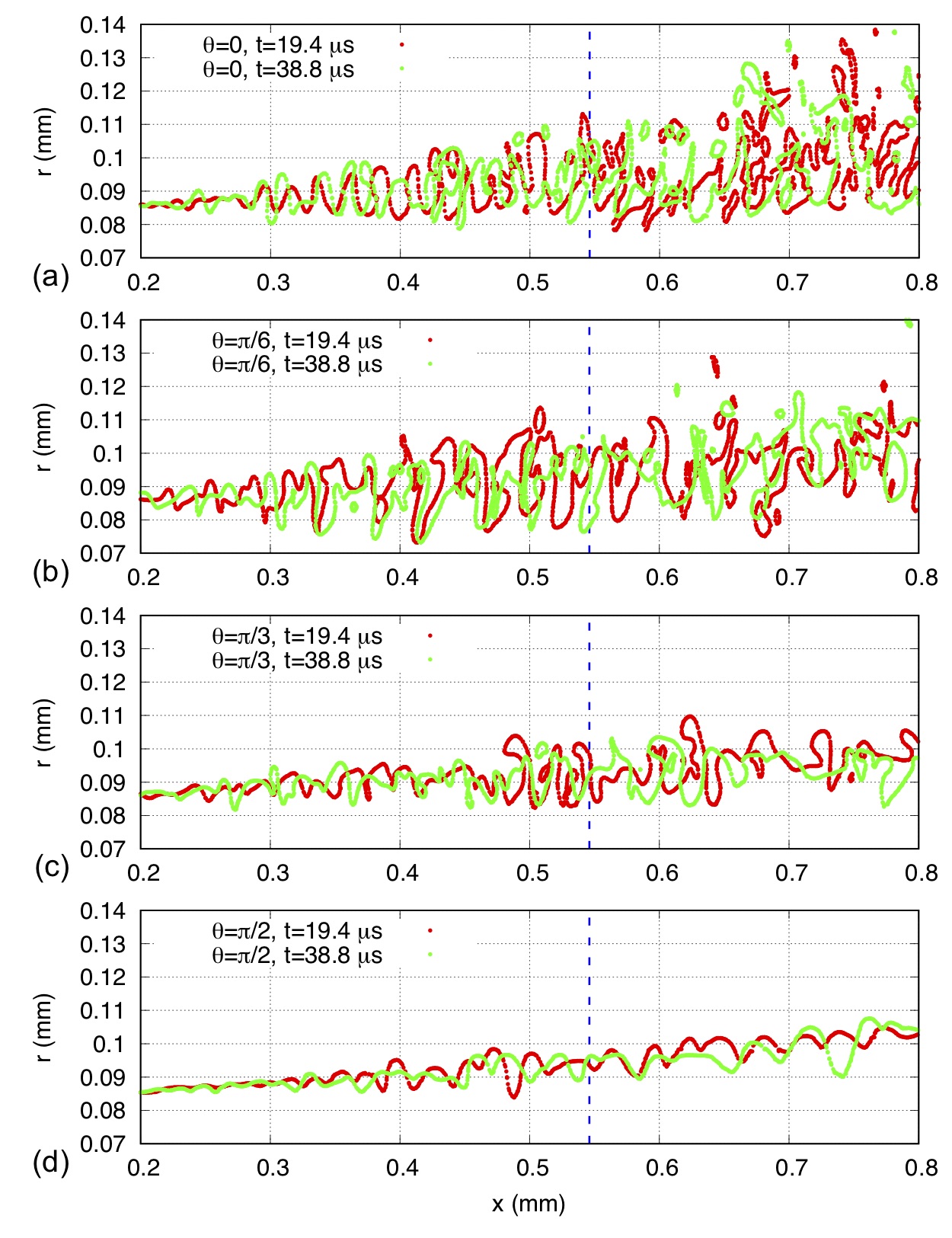}
\end{center}
\caption{Jet surface contours on planes along different azimuthal angles, (a) $\theta =$ 0, (b) $\pi/6$, (c) $\pi/3$, and (d) $\pi/2$, respectively. The blue vertical line denotes the position of the outer edge of the counterbore.  
}
\label{compare_wavelength} 
\end{figure}

In order to better show the variation of the longitudinal interfacial waves over the azimuthal angle, the jet surface contours for $\theta$ from 0 to $\pi/2$ are shown in Fig.\ \ref{compare_wavelength}. In each figure, the results for two different time instants are presented. Important wave features, such as the wavelength and amplitude, for the two different times are very similar, affirming that the portion of the jet has reached a quasi-steady state. The blue dashed lines indicate the outer boundary of the counterbore. Due to the higher liquid velocity for $\theta=0$ and $\pi/6$,  the wave amplitudes grow much faster than those for $\theta=\pi/3$ \  and $\pi/2$. The interfacial waves for small $\theta$ start to roll up and break into droplets and ligaments even within the counterbore. In the spatial region shown here, there are no droplets formed for $\theta=\pi/3$ and $\pi/2$. 
The average wavelength for $\theta=\pi/2$ is about 28 \textmu m, which is more than 45\% larger than the average wavelength for $\theta=0$. The average wave length for $\theta=0$ is only calculated for $x\lesssim0.5$ mm, as it is hard to identify individual waves after the waves roll up and break. 

\subsection{Deformation and breakup of the jet head}
\label{sec:head}
\begin{figure}[tbp]
\begin{center}
\includegraphics [width=.99\columnwidth]{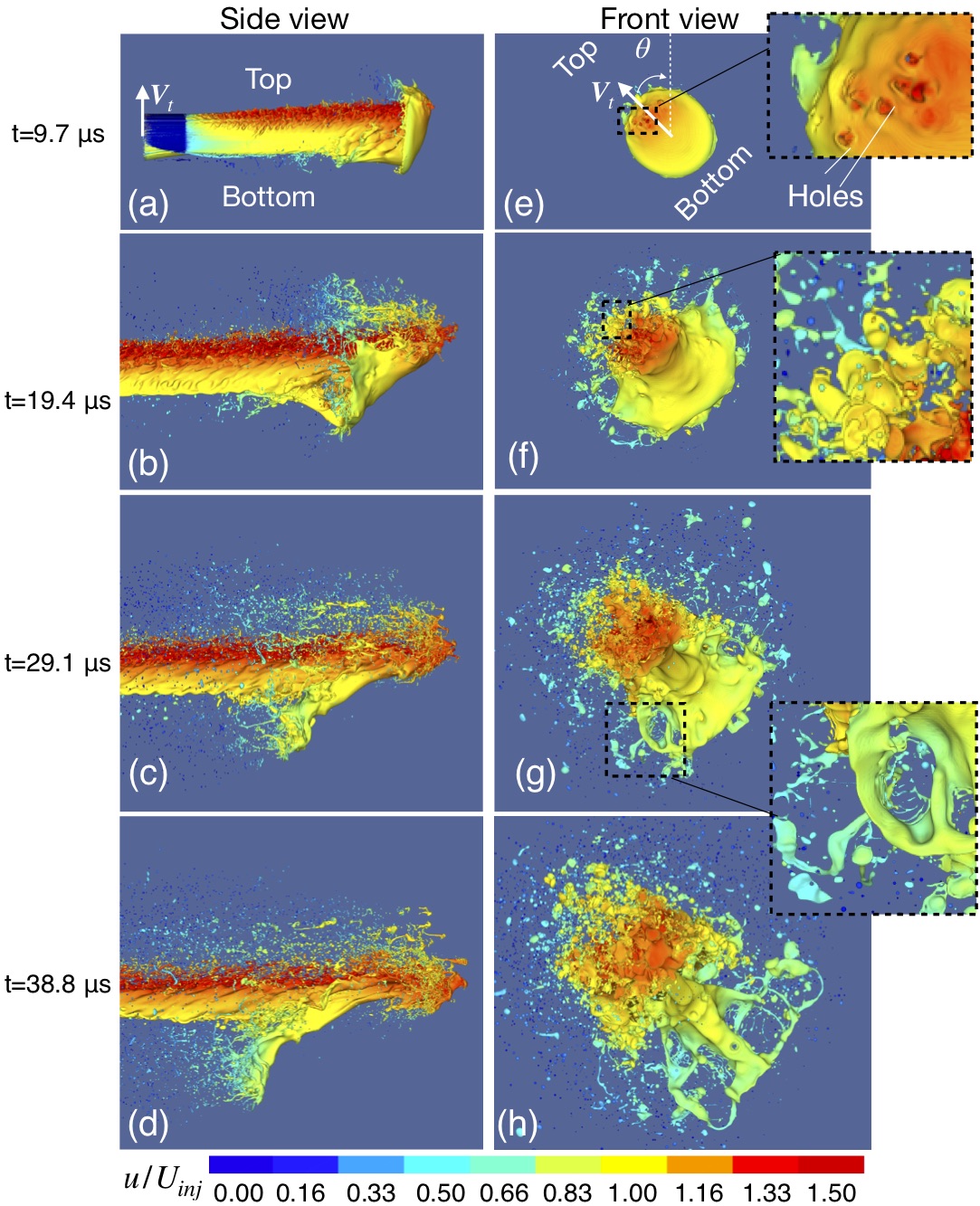}
\end{center}
\caption{Temporal evolution of the jet head from the side (a-d) and front (e-h) views. The gas-liquid interfaces are colored by the streamwise velocity. } 
\label{velocity2} 
\end{figure}

Droplets are formed not only near the jet core, but also from the continuous breakup of the jet head. Actually, the number of droplets produced due to the breakup of the jet head is significantly higher than that for the jet core.  Here, the term ``jet head" includes also the liquid sheets extended from the tip of the liquid jet. The temporal evolution of the jet head is depicted in Fig.\ \ref{velocity2}. Similar to Fig.\ \ref{velocity1}, the color represents the streamwise velocity on the interface. It can be clearly seen that the velocity at the top of the jet head is higher than that at the bottom. At early time, the shape of the head remains approximately spherical on the front view, see Fig.\ \ref{velocity2}(e). Yet as time elapses, the deformation of the jet head becomes strongly asymmetric. It can be observed from the side view that the head tilts more and more along the streamwise direction, see Figs.\ \ref{velocity2}(c) and (d). 

Due to the faster motion of the top of the jet, the liquid sheet extended from the top of the head experiences a larger aerodynamic drag. The stronger interaction with the surrounding gas results in a faster thinning of the sheets and also the earlier formation of holes in them, see the closeup of Fig.\ \ref{velocity2}(e). Holes are first observed around $|\theta|\lesssim$$\pi/6$. The holes then expand due to the Taylor-Culick rim retraction. When the holes eventually merge, the sheet breaks into small ligaments and droplets. Similar to the droplets formed near the jet core, the droplets are slowed down by the aerodynamic drag and are left behind in the wake of the jet head. 

As time elapses, the breakup of the jet head gradually extends toward the lower part. At $t=19.4$ \textmu s, the upper half of the head is almost completely broken while the bottom sheet remains relatively smooth. At $t=38.8$ \textmu s, the whole jet head is almost completely broken. The liquid velocity in the lower portion of the jet head is lower than the top. Furthermore, when the upper part of the jet head has broken, the gas can go around the head from the top, which further reduces the shear on the lower surface of the jet head. As a result, the interfacial instabilities develop slower and the breakup is less violent at the lower part of the jet head. The droplets formed from the lower part are generally larger than those from the upper part. As will be shown later, this azimuthal variation of breakup dynamics will lead to interesting asymmetric droplets statistics. 

%
%

\subsection{Turbulent vortical structures}
\begin{figure}[tbp]
\begin{center}
\includegraphics [width=.87\columnwidth]{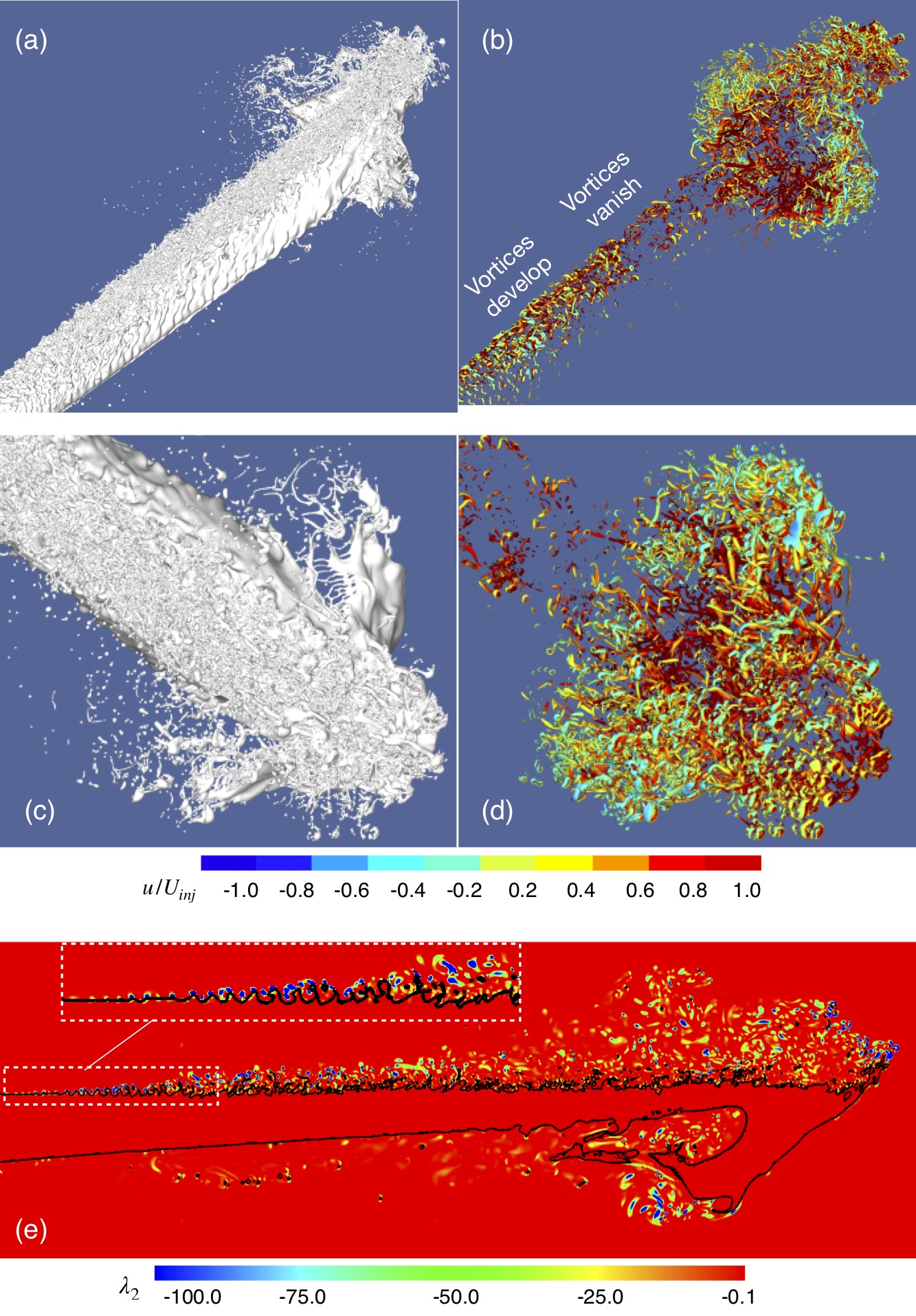}
\end{center}
\caption{Jet surfaces (a,c) and vortical structures (b,d) for $\eta=0.2$ from different views at 19.4 \textmu s. The vortices are visualized by the iso-surfaces for {$D_0\lambda_2 U_0=-100$}, colored with the streamwise velocity. (e) Contours of $\lambda_2$ on the 2D plane at $\theta=0$, with the black lines indicating the gas-liquid interfaces. }
\label{compare_lambda2} 
\end{figure}

The $\lambda_2$ criterion \cite{Jeong_1995a} is used to visualize the vortices generated around the jet, see Fig.~\ref{compare_lambda2}. 
The iso-surfaces for {$D_0\lambda_2/U_0=-100$} colored by the streamwise velocity at $t=19.4$ \textmu s are shown in Figs.\ \ref{compare_lambda2}(b) and (d) from two different views. The corresponding gas-liquid interfaces are shown in Figs.\ \ref{compare_lambda2}(a) and (c), respectively. 
The contour of $\lambda_2$ on a 2D plane along $\theta=0$ is shown in Fig.\ \ref{compare_lambda2}(e). 

Vortices are generated due to the shear instability at the interface \cite{Jarrahbashi_2014a,Zandian_2019a, Ling_2019a}. These vortices develop spatially and lead to turbulence. Due to the lower gas viscosity, the vorticity layer near the interface is significantly thinner on the gas side than that on the liquid side. As a result, the gas flow is less stable and the vortices are mainly located in the gas flow, see Fig.~\ref{compare_lambda2}(e). 

The evolution of the vortices around the jet core is closely related to the growth of the interfacial waves. Consistent with the observations in previous studies  \cite{Ling_2019a},  as the amplitudes of the interfacial waves grow spatially, more vortices are generated and the swirling strength of the vortices (characterized by the magnitude of $\lambda_2$) increases. After the interfacial waves break, the vortices gradually vanish. The number of vortices reaches its maximum at about $x/D_0=5$. Due to the stronger shear at the top of the jet,  vortices are concentrated around the upper part of the jet surface. 

A large amount of vortices are produced around the jet head, see Fig.~\ref{compare_lambda2}(d). As the gas flows over the head, vortices are formed on the upstream side of the jet head due to the shear instability, similar to those on the surfaces of the jet core. Furthermore, the gas flow separates on the downstream side of the jet head and forms a recirculation region \cite{Shinjo_2010a}. The recirculation flow itself is also unstable and becomes turbulent. Finally, when the jet head breaks into small ligaments and droplets, vortices are also produced in the wakes of these small liquid structures. 

Since the jet is progressively entering the domain, it is infeasible to perform averaging and to calculate the turbulence statistics as in previous studies of turbulent atomization \cite{Ling_2019a}. Nevertheless, the results here indicate that the turbulence near an atomizing jet is generally far from equilibrium. This non-equilibrium nature must be carefully incorporated to the sub-grid stress model if a LES simulation is to be performed. 

\subsection{Droplet statistics}
\label{sec:drop_stats}
\begin{figure}[tbp]
\begin{center}
\includegraphics [width=1.\columnwidth]{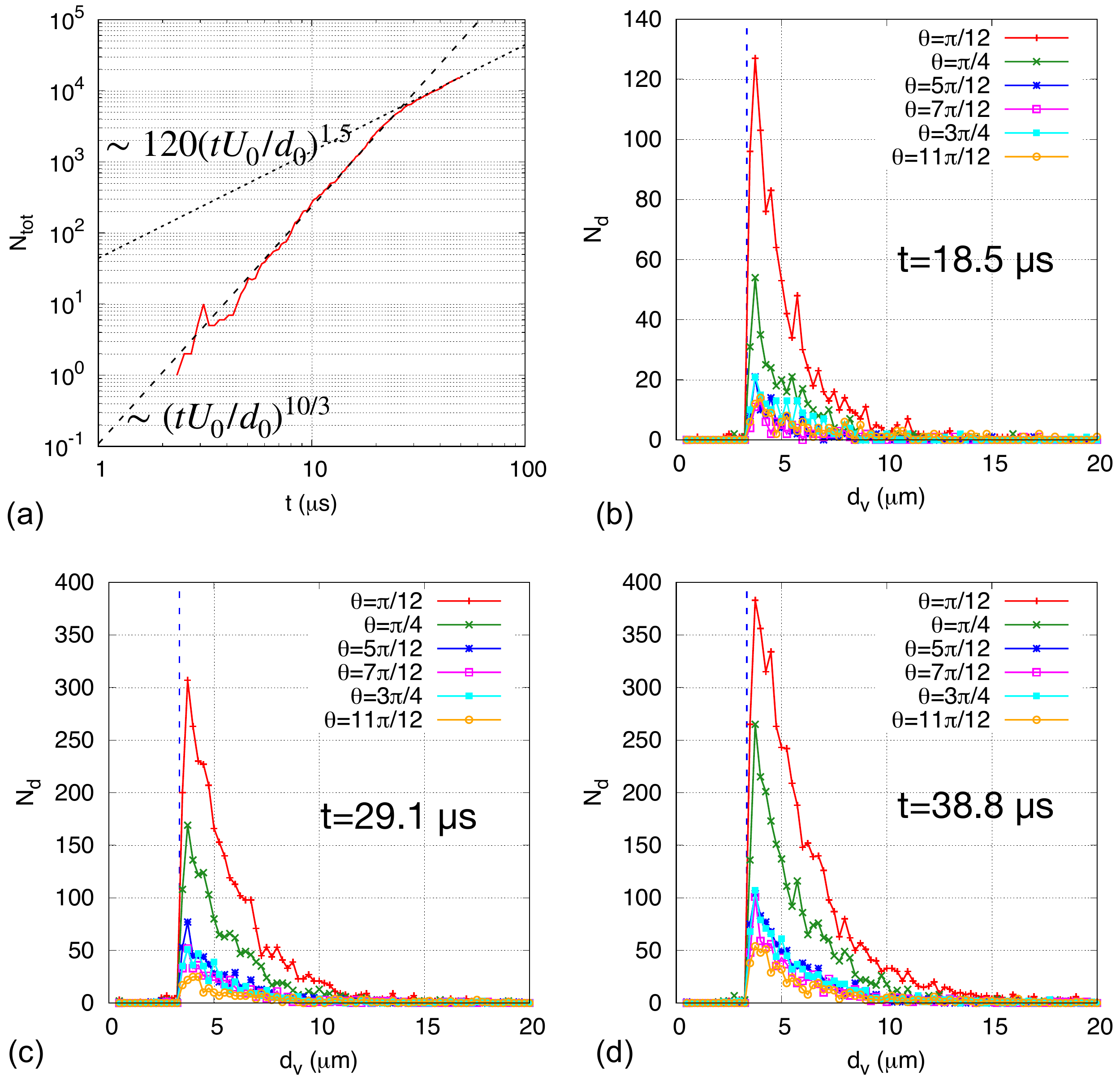}
\end{center}
\caption{Temporal evolutions of (a) the total number of droplets and (b)-(d) size distributions for different azimuthal angles. The vertical dashed lines in (b)-(d) indicate the cutoff droplet diameter $d_{v,cut}=3.35$ \textmu m.}
\label{number_statistics} 
\end{figure}

In each time snapshot of the simulation results, the individual liquid structures, such as droplets and ligaments, are identified by examining the cells with $f>0$ that are connected together. During the simulation, the droplets with a volume smaller than $(2\Delta_{x,\min})^3$ are removed, because these droplets are under resolved and removing them is helpful to stabilize the simulation. The temporal and spatial evolutions of the droplet number distributions over the volume-based droplet diameter, $d_v$,  are shown in Fig.\ \ref{number_statistics}. The vertical dashed lines in the figures indicate the cut-off droplet diameter, $d_{v,cut}$. For the L12 mesh, $d_{v,cut}=3.35$ \textmu m. 

\subsubsection{Time evolution of drop statistics}
In order to investigate the azimuthal variation of the droplet number, the droplets are counted in different azimuthal sectors $[\theta-\Delta_\theta/2,\theta+\Delta_\theta/2]$, where $\Delta_\theta$ is the span of $\theta$ for the sector. Due to the symmetry of droplet statistics with respect to the plane for $\theta=0$, the number of droplets for $\theta$ also include the droplets in the sector for $-\theta$. The number of droplets collected in the azimuthal sector centered at $\theta$ and in the diameter bin centered at $d_v$  is denoted as $N_d(t, d_v, \theta)$, which is a function of $t, d_v$, and $\theta$. Summing $N_d$ over all $\theta$ sectors and $d_v$ bins will yield the total number of droplets at a given time, $N_{tot}(t)$. The temporal evolution of $N_{tot}$ is shown in Fig.\ \ref{number_statistics} (a). As the liquid jet progressively enters the domain and breaks into droplets, $N_{tot}$  increases over time. It is interesting to notice that, the temporal growth of $N_{tot}$ exhibits two different scaling laws: at early time ($t\lesssim 27$ \textmu s) $N_{tot} \approx (tU_0/d_0)^{10/3}$, and at later time ($t\gtrsim 27$ \textmu s), $N_{tot} \approx 120(tU_0/d_0)^{1.5}$. The two scaling laws reflect the change of the breakup dynamics of the jet over time. 

As shown in section \ref{sec:head}, the breakup of the jet head first starts from its upper portion. Since the upper part of the jet head moves with larger velocity, the breakup is more violent, forming smaller droplets. As the liquid volume inflow rate is constant, the smaller droplet sizes will result in a higher rate of increase for droplet number and a faster growing power law, $N_{tot} \sim t^{10/3}$.  As time evolves, the breakup of the jet head extends toward the lower part. The breakup of the lower portion of the jet head is less intense and the droplets formed are generally larger than those formed earlier from the upper portion of the jet head. As a consequence, the rate of increase in droplet number is reduced, as reflected in the slower growing scaling law ($N_{tot} \sim t^{1.5}$).

Since a simulation snapshot contains all the droplets generated up to that time, it is difficult to identify the formation time for individual droplets. In order to investigate the statistics of droplets formed at different times, the distribution of droplet number over $d_v$ and $\theta$ at different times are shown in  Figs.\ \ref{number_statistics}(b)--(d). At $t=18.5$ \textmu s, the sector for $\theta=\pi/12$ dominates in $N_d$ and the distribution profile is relatively narrow, concentrating in the range of small $d_v$. This is consistent with the observation in Fig.\ \ref{velocity2} that the majority of the droplets earlier than $t=18.5$ \textmu s are from the breakup of the upper portion of the jet head. As a result, the droplets are located mainly at smaller $\theta$. As time evolves, the breakup of the jet head extends to larger $\theta$, and the ratio between $N_d$ for larger and smaller $\theta$ increases. Taking $d_v=4.5$ \textmu m as an example, the ratio between $N_d$ for $\theta=\pi/4$ and $\pi/12$ is around 25\% at $t=18.5$ \textmu s, and the ratio increases to about 55\% at $t=29.1$ \textmu s. Furthermore, the width of the distribution profile increases from $t=18.5$ to $t=38.8$ \textmu s. This indicates that the droplets formed at later time  biased toward larger $d_v$, which is due to the less violent breakup of the lower portion of the jet head. 

\subsubsection{Self-similar PDF for different azimuthal angles}
Another important observation can be made from Fig.\ \ref{number_statistics}, \ie, though $N_d$ varies significantly over $\theta$, the shapes of the size-distribution profiles for different $\theta$ are actually quite similar at later time ($t=29.1$ and 38.8 \textmu s). This similarity in distribution profiles for different $\theta$ can be better illustrated by the probability distribution function (PDF) $P$. The PDF of $d_v$ also depends on $\theta$ and $t$, and can be computed as
\begin{equation}
	P(d_v, \theta,t) = \frac{N(d_v,\theta,t)}{\Delta_d   \sum_d N(d_v,\theta, t)}\, ,
	\label{eq:PDF}
\end{equation}
where $\sum_d N(t,d_v,\theta)$ represents the total number droplets for $t$ and $\theta$. By definition $\int P\ \mathrm{d} d_v=1$ for all $t$ and $\theta$. 

\begin{figure}[tbp]
\begin{center}
\includegraphics [width=1.\columnwidth]{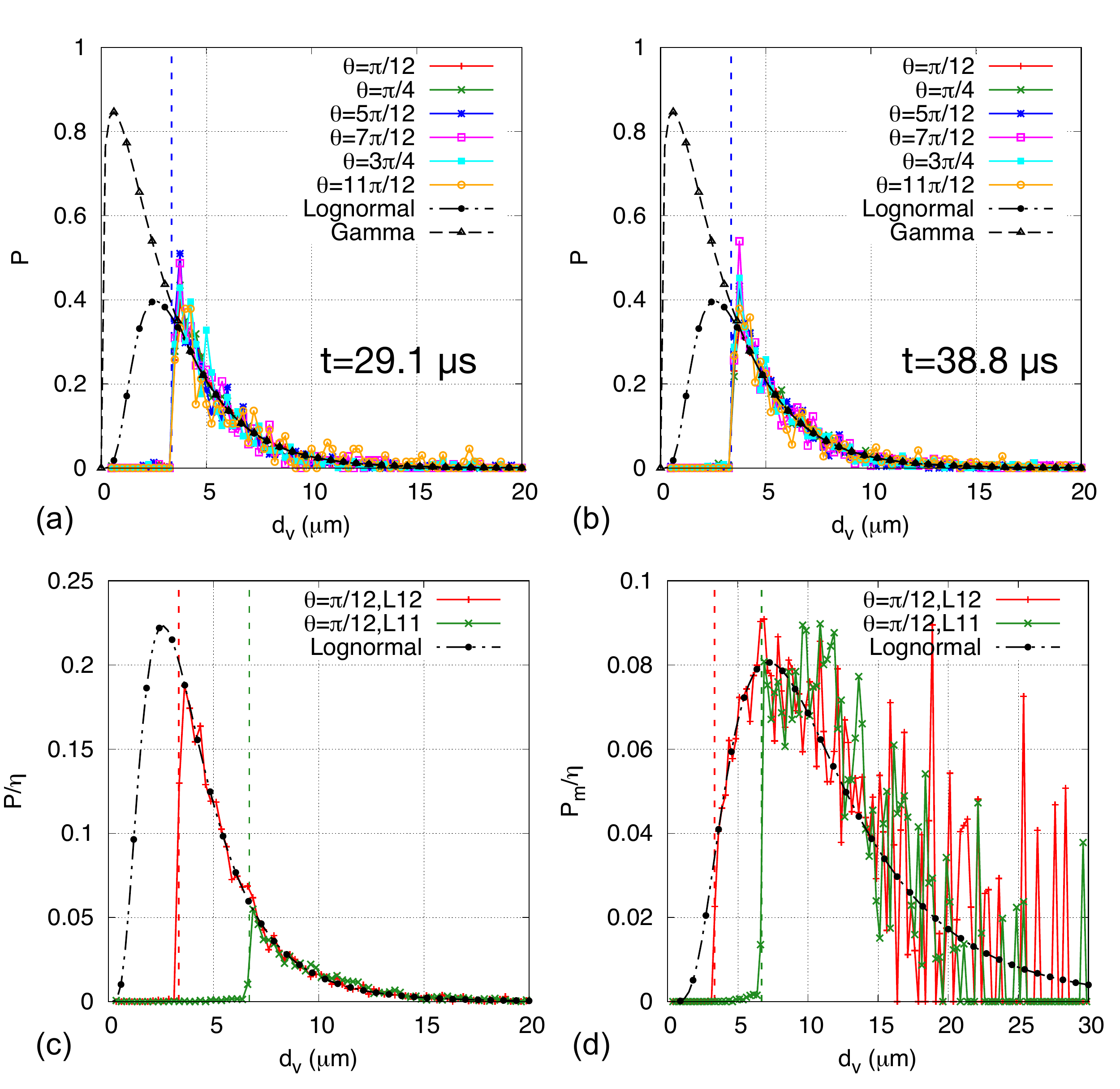}
\end{center}
\caption{Probability distribution functions (PDF) of $d_v$ for different $\theta$ at (a) $t=$29.1 and (b) 38.8 \textmu s. The lognormal and gamma functions plotted in both (a) and (b) are fitted based on the results for $\theta=\pi/12$ and $t=38.8$ \textmu s and scaled by the correction factors $\eta$. The normalized PDF ($P/\eta$) for the L11 and L12 meshes and $\theta=\pi/12$ at 38.8 \textmu s are compared with the lognormal function in (c). The droplet mass PDF of $d_v$ for $\theta=\pi/12$ and $t=38.8$ \textmu s and the L11 and L12 meshes are shown in (d). The vertical dashed lines indicate the cut-off droplet diameter $d_{v,cut}$ for the corresponding mesh. }
\label{drop_statistics} 
\end{figure}

It can be observed from Figs.\ \ref{drop_statistics}(a) and (b) that the profiles of $P$ for different $\theta$ tend to collapse for both $t=29.1$ and  38.8 \textmu s. In other words, although the droplet number $N_d$ varies significantly over $\theta$, the PDF $P$ does not. Furthermore, the collapsed profile of $P$ varies little over time. As a result, $P$ at later time can be approximated by a self-similar form, $P_{sim}$, namely
\begin{equation}
	P(d_v,\theta,t) \approx P_{sim}(d_v)\, ,
\end{equation}
while $P_{sim}$ is only a function of $d_v$ and does not depend on $t$ and $\theta$. 

\subsubsection{Estimate for the statistics of under-resolved droplets}
It can be observed from Fig.\ \ref{drop_statistics} that, the peaks of $P$ are right next to $d_{v,cut}$, which seems to indicate that there exist droplets that are under resolved ($d_v<d_{v,cut}$) in the present simulation. In order to estimate the statistics of these under-resolved droplets, the model distribution functions, including the lognormal and gamma distribution functions, are employed to fit the PDF for resolved droplets ($d_v>d_{v,cut}$). The expressions for the lognormal and gamma distributions are given as 
\begin{align}
	P_L(d_\text{v}) = \frac{\eta}{d_{\text{v}} \hat{\sigma} \sqrt{2\pi}} \exp \bigg[-\frac{(\ln d_{\text{v}}-\hat{\mu})^2}{2\hat{\sigma}^2} \bigg]  \, ,
	\label{eq:log_func}
\end{align}
where $\hat{\mu}$ and $\hat{\sigma}^2$ are the mean and variance of $\ln d_{\text{v}}$, 
and 
\begin{align}
	P_G(d_\text{v}) = \eta \frac{\hat{\beta}^{\hat{\alpha}}}{\Gamma(\hat{\alpha})} d_\text{v}^{\hat{\alpha}-1} \exp (-\hat{\beta} d_\text{v})  \,   
	\label{eq:gamma_func}
\end{align}
where $\hat{\alpha} = (\tilde{\mu}/\tilde{\sigma})^2$ and $\hat{\beta} = \hat{\alpha}/\tilde{\mu}$ with $\tilde{\mu}$ and $\tilde{\sigma}^2$ the mean and the variance of $d_\text{v}$, respectively. The correction factor $\eta$ is introduced to account for the under-resolved droplets. The lognormal and gamma profiles plotted in Figs.\ \ref{drop_statistics}(a) and (b) are based on the results for $d_v \in [4:20]$ \textmu m and $\theta= \pi/12$ at $t=38.8$ \textmu s. The fitting parameters are $(\hat{\mu},\hat{\sigma}) = (1.29,0.58)$ and  $(\hat{\alpha},\hat{\beta}) =  (1.26,0.44)$ for the lognormal and gamma functions, respectively. It can be observed that the fitted profiles agree well with results of $P$ for different $t$ and $\theta$. 

The correction factors for the lognormal and gamma distributions are $\eta=$1.8 and 3.2, respectively. If we assume that the PDF for the droplets generated followed the lognormal or gamma distributions, the percentages of the under-resolved droplets in terms of number are about $(\eta-1)/\eta$=44\% and 69\%, respectively. Previous numerical and experimental studies have shown that the lognormal function fits better the gradual decay of $P$ for larger $d_v$ \cite{Sotolongo-Costa_1996a, Ling_2017a}. The diameter at the peak of $P$ estimated by the lognormal function is about $d_v=$2.6 \textmu m, which is about twice of $\Delta_{x,\min}$ and is slightly smaller than $d_{v,cut}=3.35$ \textmu m for the L12 mesh. 

The results for the normalized PDF, namely $P/\eta$, are shown in Fig.\ \ref{drop_statistics}(c). The simulation results for the L11 and L12 meshes ($\theta= \pi/12$ and $t=38.8$ \textmu s) are compared with the lognormal function. The integration of the normalized lognormal function $\int_0^\infty (P_L/\eta) \mathrm{d}d_v =1$. The correction factor $\eta$ for the L11 mesh results is about 6.5. In other words, when the coarser L11 mesh is used, the percentage of under-resolved droplets increases to about 85\%. Nevertheless, it is observed that the normalized PDF for the resolved droplets for the L11 mesh agrees well with the PDF for the L12 mesh and also the lognormal function. This seems to indicate that the statistics of the droplets is not influenced by leaving some small droplets under resolved, assuming that the important primary breakup processes (such as the interfacial waves and the jet head breakups) are reasonably captured. 

Furthermore, the percentage of under-resolved droplets may seem to be high in terms of number, but actually they take only a small portion of the total mass (or volume) of the droplets formed. The droplet mass PDF of $d_v$, $P_m$, is defined as 
\begin{equation}
	P_m(d_v, t, \theta) = \frac{m(d_v,\theta, t)}{\Delta_d  \sum_d m(d_v,\theta,t)} 
\end{equation}
where $m(d_v,\theta, t)$ denotes the total mass of droplets for $d_v,\theta$ and $t$. Since the droplet fluid density is taken to be constant, so $P_m$ can be related to $P$ as
\begin{equation}
	P_m(d_v, t, \theta) =  \frac{N(d_v,\theta, t) d_v^3}{\Delta_d  \sum_d [N(d_v,\theta,t)d_v^3]} = P_d \frac{d_v^3}{\langle d_v^3 \rangle}
\end{equation}
where
\begin{equation}
	\langle d_v^3 \rangle = \int_0^\infty P(d_v) d_v^3 \mathrm{d}d_v
\end{equation}
 is the mean of $d_v^3$ and it is computed that  $\langle d_v^3 \rangle=$220.57 \textmu m$^3$ according to the fitted lognormal function.
The results of $P_m$ for $\theta= \pi/12$, $t=38.8$  \textmu s, and the L11 and L12 meshes  are shown in Fig.\ \ref{drop_statistics}(d).
The simulation results for $P_m$ are more noisy due to the factor of $d_v^3$. The peak of $P_m$ can be identified at about $d_v=7$ \textmu m, which is about the $d_{v,cut}$ for the L11 mesh and is about twice the $d_{v,cut}$ (about four times of $\Delta_{x,min}$) for the L12 mesh. More important, it can be computed from the lognormal fit that the percentage of the under-resolved droplets in terms of droplet mass for the L12 mesh is about 3.1\%, which is actually quite small. Therefore, the present simulation with the finer L12 mesh does capture the majority of droplets in terms of mass or volume. 

\subsubsection{PDF for azimuthal angle}
\begin{figure}[tbp]
\begin{center}
\includegraphics [width=.55\columnwidth]{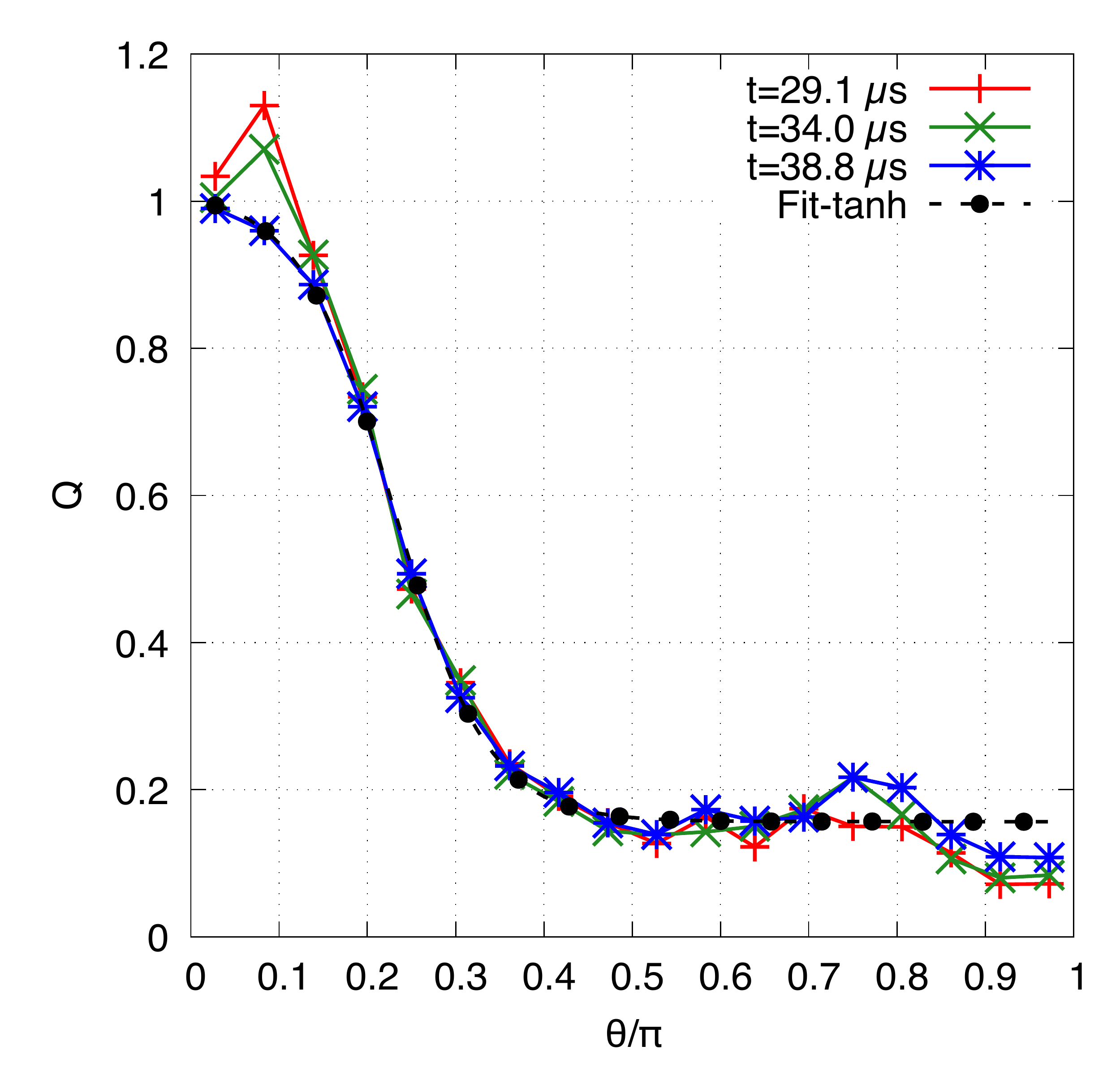}
\end{center}
\caption{The PDF of droplet number for the azimuthal angle $\theta$ at different times. The fitted function is a hyperbolic tangent function. 
}
\label{drop_statistics_azimu} 
\end{figure}

The PDF of the azimuthal angle $\theta$ is defined as
\begin{equation}
	Q(\theta,t) = \frac{ \sum_d N(d_v,\theta, t)}{\Delta_{\theta}  N_{tot}(t)}\, ,
\end{equation}
which is a function of $\theta$ and $t$. 
It can be shown that $\int_0^\pi Q \mathrm{d}\theta = 1$ for all $t$.
The results for $Q$ at different times are plotted in Fig.\ \ref{drop_statistics_azimu}. Similar to $P$, it is observed that $Q$ varies only slightly over time for $t\gtrsim29.1$ \textmu s, so we can approximate $Q$ with a similar profile that depends on $\theta$ only
\begin{align}
	Q(t,\theta) \approx Q_{sim}(\theta)\, .
\end{align}
The variation of $Q$ over $\theta$ reflects the asymmetric breakup dynamics of the jet head and the jet core. 
It is worth noting that the droplets have a small azimuthal velocity when they are just generated, so the change of droplet location in the $\theta$ coordinate is generally small. The hyperbolic tangent function well captures the decrease of $Q_{sim}$ over $\theta$ between $0$ and $\pi/2$. There exist mild variations of $Q_{sim}$ between $\theta=\pi/2$ and $\pi$, but the amplitudes of those variations are much smaller than the change from $\theta=0$ to $\pi/2$. 
The hyperbolic tangent function fitted based on the data at $t=$38.8 \textmu s is given as
\begin{equation}
	Q_{sim}(\theta) \approx 0.0429 \tanh [-9.29(\theta/\pi-0.229)] + 0.585\,,
	\label{eq:Q_fit}
\end{equation} 
which is plotted in Fig.\ \ref{drop_statistics_azimu} and is shown to be a good approximation of $Q$. 

\subsubsection{Model to estimate droplet number}
\begin{figure}[tbp]
\begin{center}
\includegraphics [width=.55\columnwidth]{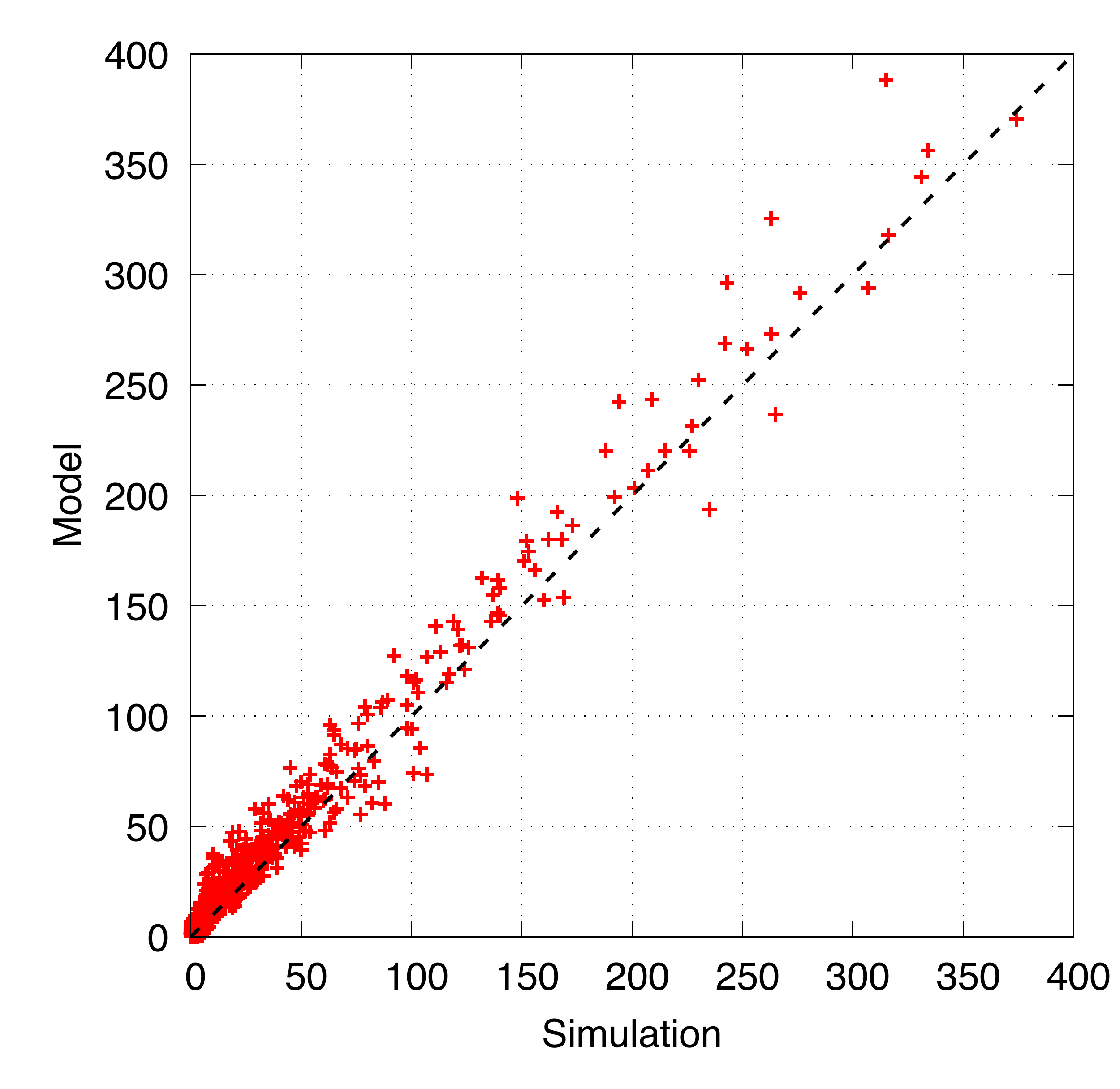}
\end{center}
\caption{The number of droplets estimated by the model (Eq.\ \eqr{drop_num_est}) are compared with the simulation results. The range of droplet size is $3.5<d_v<30$ \textmu m, and the bin width is $\Delta_d=0.25$ \textmu m. The angle of the azimuthal sector $\Delta_\theta=\pi/6$. The data plotted include three time snapshots at  $t=29.1$, $39.5$ and $38.8$ \textmu s. 
}
\label{fig:drop_num_estimate} 
\end{figure}

Finally, the results obtained previously for (1) the time scaling law for the total number of droplets $N_{tot}(t)$ at later time, \ie, $N_{tot} \approx 120 (tU_0/d_j)^{1.5}$, (2) the self-similar PDF of droplet diameter, $P_{sim}(d_v)$, which is approximated by the lognormal function $P_L(d_v)$ (Eq.\ \eqr{log_func} with $(\eta, \hat{\mu},\hat{\sigma}) = (1.8,1.29,0.58)$), and (3) the self-similar PDF for the azimuthal angle, $Q_{sim}(\theta)$ (Eq.\ \eqr{Q_fit}), lead to a useful model to estimate the number of droplets in any droplet size bin and azimuthal angle sector at later time of the primary breakup ($t\gtrsim 27$ \textmu s): 
\begin{equation}
	N_{est}(t,d_v,\theta) =  N_{tot}(t) Q_{sim}(\theta) P_{sim}(d_v) \Delta_\theta  \Delta_d.
	\label{eq:drop_num_est}
\end{equation}
The droplet numbers for different $d_v, \theta$, and $t$ estimated by the model (Eq.\ \eqr{drop_num_est}) are compared with the simulation results in Fig.\ \ref{fig:drop_num_estimate}. The data plotted here include three time snapshots at  $t=29.1$, $39.5$ and $38.8$ \textmu s for the droplet diameter range from 3.5 to 30 \textmu m. The bin width for $d_v$ is $\Delta_d=0.25$ \textmu m and the angle of the azimuthal sector $\Delta_\theta=\pi/6$. It is clearly shown that the model yields good estimates to the simulation results. The model exhibited a simple explicit form and accurately captures the droplets number distribution over $d_v, \theta$, and $t$, therefore, it is very useful in practical applications. For example, the model can be applied to specify the conditions of droplets at the inlet in a Lagrangian spray simulation where the primary breakup process is not directly simulated. 

\section{Conclusions}
\label{sec:conclusions}
The primary breakup of a gasoline surrogate jet is investigated through detailed numerical simulation. The interfacial two-phase flow is resolved using the \emph{Basilisk} solver with a momentum-conserving volume-of-fluid method. The injection conditions are similar to the Engine Combustion Network (ECN) Spray G operating conditions. To focus the computational resources on resolving the liquid jet, the injector geometry is simplified. The effect of the internal flow in the injector on the jet dynamics is modeled through a nonzero injection angle specified at the inlet. A parametric study is performed for the injection angle. The simulation results for different injection angles are compared with the experimental measurements for the jet penetration length and the jet deflection angle to identify the injection angle ($\eta=\tan \alpha =0.2$) that best represents the Spray G conditions. The effects of the inlet boundary condition, numerical method, and mesh resolution are systematically investigated, affirming that the simulation approach is effective in resolving both the macro-scale and micro-scale breakup features. The nonzero injection angle introduces an azimuthally varying velocity within the liquid jet. As a consequence of that, the shear-induced interfacial waves on the jet core and the formation of liquid lobes and fingers become strongly asymmetric: the wavelengths for the longitudinal waves on the top of the jet are significantly smaller than those on the lateral sides. The deformation and breakup of the jet head are also influenced by the non-uniform velocity. Since the upper portion of the jet head moves faster than the lower portion, the jet head tilts in the streamwise direction and furthermore, the upper portion breaks earlier and more violently than the lower portion. This time-dependent and asymmetric breakup dynamics of the jet head results in two different scaling laws for the total droplet number at the early and later times. While the former scaling law corresponds to the smaller droplets generated from the earlier and more violent breakup of the upper portion of the jet head, the latter is dictated by larger droplets produced by the later breakup of the lower portion of the jet head. The distribution of the droplet number over the volume-based droplet diameter is presented as a function of time and azimuthal angle $\theta$. Though the droplet-number distribution varies significantly over $\theta$, the probability density functions (PDF)  for different $\theta$ collapse to a self-similar profile. The self-similar PDF is fitted with both the lognormal and gamma distribution functions. The results for PDF suggest that there exist droplets that are smaller than the cut-off droplet diameter (droplet volume smaller than $(2\Delta_{x,min})^3$) and thus are under resolved in the present simulation. The PDF for the resolved droplets for the L11 and L12 meshes agree well with the lognormal function, indicating that the size-distribution of resolved droplets are not influenced by leaving some tiny droplets under resolved, assuming  the mesh resolution is fine enough to capture the important micro-scale breakup features like the interfacial waves and the jet head deformation. The percentage and statistics of the tiny under-resolved droplets are estimated through the lognormal function. It is shown that about 3.1\% of the total droplet mass are under resolved by the L12 mesh. The PDF of the azimuthal angle is also presented. The decrease of PDF over the azimuthal angle is well represented by a hyperbolic tangent function. Both the PDF of $d_v$ and $\theta$ vary little over time at later time ($t\gtrsim 27$ \textmu s). Based on these self-similar PDF, a model has been proposed to predict the droplet number for an arbitrary droplet diameter and azimuthal angle at later time of the primary breakup. The model predictions are shown to agree well with the simulation results. 

{The present study has only simulated for a short physical time, compared to the whole injection duration of the spray G operation conditions. Therefore, the atomizing jet in the computation domain has not reached a statistically stationary state. To measure time-average two-phase turbulent flow properties, the simulation needs to be run for a much longer time (twice or even more). Such a simulation will be relegated to the future work. }

\section*{Acknowledgements}
This research has been supported by the National Science Foundation (NSF \#1853193). The authors also acknowledge the Extreme Science and Engineering Discovery Environment (XSEDE)  and the Texas Advanced Computing Center (TACC) for providing the computational resources that have contributed to the research results reported in this paper. We would also thank Dr.\ Christopher Powell and Dr.\ Alan Kastengren at Argonne National Lab for the helpful communication on the details of their experiment.

\section*{References}

\begin{thebibliography}{10}
\expandafter\ifx\csname url\endcsname\relax
  \def\url#1{\texttt{#1}}\fi
\expandafter\ifx\csname urlprefix\endcsname\relax\def\urlprefix{URL }\fi
\expandafter\ifx\csname href\endcsname\relax
  \def\href#1#2{#2} \def\path#1{#1}\fi

\bibitem{Zhao_1999a}
F.~Zhao, M.-C. Lai, D.~L. Harrington, Automotive spark-ignited direct-injection
  gasoline engines, Prog.~Energ.~Combust.~Sci. 25 (1999) 437--562.

\bibitem{Mitroglou_2006a}
N.~Mitroglou, J.~M. Nouri, M.~Gavaises, C.~Arcoumanis, Spray characteristics of
  a multi-hole injector for direct-injection gasoline engines, Int.~J.~Engine
  Res. 7~(3) (2006) 255--270.

\bibitem{Wang_2015a}
Z.~Wang, A.~Swantek, R.~Scarcelli, D.~Duke, A.~Kastengren, C.~F. Powell,
  S.~Som, R.~Reese, K.~Freeman, Y.~Zhu, {LES} of diesel and gasoline sprays
  with validation against {X-ray} radiography data, SAE Int.~J.~Fuels Lubr. 8
  (2015) 147--159.

\bibitem{Duke_2017a}
D.~J. Duke, A.~L. Kastengren, K.~E. Matusik, A.~B. Swantek, C.~F. Powell,
  R.~Payri, D.~Vaquerizo, L.~Itani, G.~Bruneaux, R.~O. Grover~Jr, Internal and
  near nozzle measurements of engine combustion network ``{Spray G}'' gasoline
  direct injectors, Exp.~Therm.~Fluid Sci. 88 (2017) 608--621.

\bibitem{Khan_2017a}
M.~M. Khan, J.~Helie, M.~Gorokhovski, N.~A. Sheikh, Air entrainment in high
  pressure multihole gasoline direct injection sprays, J.~Appl.~Fluid Mech. 10
  (2017) 1223--1234.

\bibitem{Sphicas_2018a}
P.~Sphicas, L.~M. Pickett, S.~A. Skeen, J.~H. Frank, Inter-plume aerodynamics
  for gasoline spray collapse, Int.~J.~Engine Res. 19~(10) (2018) 1048--1067.

\bibitem{Payri_2017a}
R.~Payri, F.~J. Salvador, P.~Marti-Aldaravi, D.~Vaquerizo, {ECN} {Spray G}
  external spray visualization and spray collapse description through
  penetration and morphology analysis, Appl.~Therm.~Eng. 112 (2017) 304--316.

\bibitem{Reitz_1982a}
R.~D. Reitz, F.~V. Bracco, Mechanism of atomization of a liquid jet,
  Phys.~Fluids 25 (1982) 1730--1742.

\bibitem{Lin_1998a}
S.~P. Lin, R.~D. Reitz, Drop and spray formation from a liquid jet,
  Annu.~Rev.~Fluid Mech. 30 (1998) 85--105.

\bibitem{Aleiferis_2010a}
P.~G. Aleiferis, J.~Serras-Pereira, Z.~Van~Romunde, J.~Caine, M.~Wirth,
  Mechanisms of spray formation and combustion from a multi-hole injector with
  {E85} and gasoline, Combust.~Flame 157 (2010) 735--756.

\bibitem{Payri_2016a}
R.~Payri, J.~Gimeno, P.~Marti-Aldaravi, D.~Vaquerizo, Internal flow
  characterization on an ecn gdi injector, Atomization Spray 26 (2016)
  889--919.

\bibitem{Agarwal_2020a}
A.~Agarwal, M.~F. Trujillo, The effect of nozzle internal flow on spray
  atomization, Int.~J.~Engine Res. 21 (2020) 55--72.

\bibitem{Heindel_2018a}
T.~Heindel, X-ray imaging techniques to quantify spray characteristics in the
  near field, Atomization Spray 28 (2018) 1029--1059.

\bibitem{Gorokhovski_2008a}
M.~Gorokhovski, M.~Herrmann, Modeling primary atomization, Annu.~Rev.~Fluid
  Mech. 40 (2008) 343--366.

\bibitem{Fuster_2009a}
D.~Fuster, A.~Bagu{\'e}, T.~Boeck, L.~Le~Moyne, A.~Leboissetier, S.~Popinet,
  P.~Ray, R.~Scardovelli, S.~Zaleski, Simulation of primary atomization with an
  octree adaptive mesh refinement and {VOF} method, Int.~J.~Multiphase Flow 35
  (2009) 550--565.

\bibitem{Lebas_2009a}
R.~Lebas, T.~Menard, P.~A. Beau, A.~Berlemont, F.-X. Demoulin, Numerical
  simulation of primary break-up and atomization: {DNS} and modelling study,
  Int.~J.~Multiphase Flow 35 (2009) 247--260.

\bibitem{Desjardins_2010a}
O.~Desjardins, H.~Pitsch, Detailed numerical investigation of turbulent
  atomization of liquid jets, Atomization Spray 20 (2010) 311---336.

\bibitem{Shinjo_2010a}
J.~Shinjo, A.~Umemura, Simulation of liquid jet primary breakup: Dynamics of
  ligament and droplet formation, Int.~J.~Multiphase Flow 36 (2010) 513--532.

\bibitem{Li_2016a}
X.~Li, M.~C. Soteriou, High fidelity simulation and analysis of liquid jet
  atomization in a gaseous crossflow at intermediate weber numbers,
  Phys.~Fluids 28 (2016) 082101.

\bibitem{Ling_2017a}
Y.~Ling, D.~Fuster, S.~Zaleski, G.~Tryggvason, Spray formation in a quasiplanar
  gas-liquid mixing layer at moderate density ratios: A numerical closeup,
  Phys.~Rev.~Fluids 2 (2017) 014005.

\bibitem{Shao_2017a}
C.~Shao, K.~Luo, Y.~Yang, J.~Fan, Detailed numerical simulation of swirling
  primary atomization using a mass conservative level set method,
  Int.~J.~Multiphase Flow 89 (2017) 57--68.

\bibitem{Ling_2019a}
Y.~Ling, D.~Fuster, G.~Tryggvasson, S.~Zaleski, A two-phase mixing layer
  between parallel gas and liquid streams: multiphase turbulence statistics and
  influence of interfacial instability, J.~Fluid Mech. 859 (2019) 268--307.

\bibitem{Hasslberger_2019a}
J.~Hasslberger, S.~Ketterl, M.~Klein, N.~Chakraborty, Flow topologies in
  primary atomization of liquid jets: a direct numerical simulation analysis,
  J.~Fluid Mech. 859 (2019) 819--838.

\bibitem{Lakehal_2012a}
D.~Lakehal, M.~Labois, C.~Narayanan, Advances in the {Large-Eddy} and interface
  simulation ({LEIS}) of interfacial multiphase flows in pipes,
  Prog.~Comput.~Fluid Dyn. 12 (2012) 153--163.

\bibitem{Agbaglah_2017a}
G.~Agbaglah, R.~Chiodi, O.~Desjardins, Numerical simulation of the initial
  destabilization of an air-blasted liquid layer, J. Fluid Mech. 812 (2017)
  1024--1038.

\bibitem{Aniszewski_2016a}
W.~Aniszewski, Improvements, testing and development of the {ADM}-$\tau$
  sub-grid surface tension model for two-phase {LES}, J.~Comput.~Phys. 327
  (2016) 389--415.

\bibitem{Dukowicz_1980a}
J.~K. Dukowicz, A particle-fluid numerical model for liquid sprays,
  J.~Comput.~Phys. 35 (1980) 229--253.

\bibitem{Hoyas_2013a}
S.~Hoyas, A.~Gil, X.~Margot, D.~Khuong-Anh, F.~Ravet, Evaluation of the
  eulerian--lagrangian spray atomization (elsa) model in spray simulations: 2d
  cases, Math.~Comput.~Model. 57 (2013) 1686--1693.

\bibitem{Aguerre_2019a}
H.~J. Aguerre, N.~M. Nigro, Implementation and validation of a lagrangian spray
  model using experimental data of the {ECN} {Spray G} injector, Comput.~Fluids
  190 (2019) 30--48.

\bibitem{Paredi_2020a}
D.~Paredi, T.~Lucchini, G.~{D'Errico}, A.~Onorati, L.~Pickett, J.~Lacey,
  Validation of a comprehensive computational fluid dynamics methodology to
  predict the direct injection process of gasoline sprays using {Spray G}
  experimental data, Int.~J.~Engine Res. 21~(1) (2020) 199--216.

\bibitem{Beale_1999a}
J.~C. Beale, R.~D. Reitz, Modeling spray atomization with the
  {Kelvin-Helmholtz/Rayleigh-Taylor} hybrid model, Atomization Spray 9 (1999)
  623--650.

\bibitem{Duret_2013a}
B.~Duret, J.~Reveillon, T.~Menard, F.~X. Demoulin, Improving primary
  atomization modeling through {DNS} of two-phase flows, Int.~J.~Multiphase
  Flow 55 (2013) 130--137.

\bibitem{Vallet_1999a}
A.~Vallet, R.~Borghi, An {Eulerian} model of atomization of a liquid jet, C. R.
  Acad. Sci. Paris, s{\'e}rie {II b} 327 (1999) 1015--1020.

\bibitem{Sparacino_2019a}
S.~Sparacino, F.~Berni, A.~d'Adamo, V.~K. Krastev, A.~Cavicchi, L.~Postrioti,
  Impact of the primary break-up strategy on the morphology of {GDI} sprays in
  {3D}-{CFD} simulations of multi-hole injectors, Energies 12 (2019) 2890.

\bibitem{Maxey_1983a}
M.~R. Maxey, J.~J. Riley, Equation of motion for a small rigid sphere in a
  nonuniform flow, Phys.~Fluids 26 (1983) 883--889.

\bibitem{Michaelides_1994a}
E.~E. Michaelides, Z.~Feng, Heat transfer from a rigid sphere in a nonuniform
  flow and temperature field, Int.~J.~Heat Mass Transfer 37 (1994) 2069--2076.

\bibitem{Ling_2016a}
Y.~Ling, S.~Balachandar, M.~Parmar, Inter-phase heat transfer and energy
  coupling in turbulent dispersed multiphase flows, Phys.~Fluids 28 (2016)
  033304.

\bibitem{Piazzullo_2018a}
D.~Piazzullo, M.~Costa, L.~Allocca, A.~Montanaro, V.~Rocco, Schlieren and {Mie}
  scattering techniques for the {ECN} ``{spray G}'' characterization and 3d cfd
  model validation, Int.~J.~Numer.~Methods Heat Fluid Flow 28 (2018) 498--515.

\bibitem{Sphicas_2017a}
P.~Sphicas, L.~M. Pickett, S.~Skeen, J.~Frank, T.~Lucchini, D.~Sinoir,
  G.~{D'Errico}, K.~Saha, S.~Som, A comparison of experimental and modeled
  velocity in gasoline direct-injection sprays with plume interaction and
  collapse, SAE Int.~J.~Fuels Lubr. 10 (2017) 184--201.

\bibitem{Di-Ilio_2019a}
G.~Di-Ilio, V.~K. Krastev, G.~Falcucci, Evaluation of a scale-resolving
  methodology for the multidimensional simulation of {GDI} sprays, Energies 12
  (2019) 2699.

\bibitem{Navarro-Martinez_2020a}
S.~Navarro-Martinez, G.~Tretola, M.~R. Yosri, R.~L. Gordon, K.~Vogiatzaki, An
  investigation on the impact of small-scale models in gasoline direct
  injection sprays ({ECN} {Spray G}), Int.~J.~Engine Res. 21 (2020) 217--225.

\bibitem{Befrui_2016a}
B.~Befrui, A.~Aye, A.~Bossi, L.~E. Markle, D.~L. Varble, {ECN} {GDI} {Spray G}:
  Coupled {LES} jet primary breakup-{Lagrangian} spray simulation and
  comparison with data, in: ILASS Americas 28th Annual Conference, 2016.

\bibitem{Yue_2020a}
Z.~Yue, M.~Battistoni, S.~Som, Spray characterization for engine combustion
  network {Spray G} injector using high-fidelity simulation with detailed
  injector geometry, Int.~J.~Engine Res. 21 (2020) 226--238.

\bibitem{Fuster_2018a}
D.~Fuster, S.~Popinet, An all-mach method for the simulation of bubble dynamics
  problems in the presence of surface tension, J.~Comput.~Phys. 374 (2018)
  752--768.

\bibitem{Weymouth_2010a}
G.~D. Weymouth, D.~K.-P. Yue, Conservative volume-of-fluid method for
  free-surface simulations on cartesian-grids, J.~Comput.~Phys. 229~(8) (2010)
  2853--2865.

\bibitem{Scardovelli_1999a}
R.~Scardovelli, S.~Zaleski, Direct numerical simulation of free-surface and
  interfacial flow, Annu.~Rev.~Fluid Mech. 31 (1999) 567--603.

\bibitem{Aulisa_2007a}
E.~Aulisa, S.~Manservisi, R.~Scardovelli, S.~Zaleski, Interface reconstruction
  with least-squares fit and split advection in three-dimensional cartesian
  geometry, J.~Comput.~Phys. 225 (2007) 2301--2319.

\bibitem{Scardovelli_2000a}
R.~Scardovelli, S.~Zaleski, Analytical relations connecting linear interfaces
  and volume fractions in rectangular grids, J.~Comput.~Phys. 164 (2000)
  228--237.

\bibitem{Vaudor_2017a}
G.~Vaudor, T.~M\'enard, W.~Aniszewski, M.~Doring, A.~Berlemont, A consistent
  mass and momentum flux computation method for two phase flows. {Application}
  to atomization process, Comput.~Fluids 152 (2017) 204--216.

\bibitem{Fuster_2019a}
D.~Fuster, T.~Arrufat, M.~Crialesi-Esposito, Y.~Ling, L.~Malan, S.~Pal,
  R.~Scardovelli, G.~Tryggvason, S.~Zaleski, A momentum-conserving, consistent,
  volume-of-fluid method for incompressible flow on staggered grids,
  arXiv:1811.12327 (2019).

\bibitem{Francois_2006a}
M.~M. Francois, S.~J. Cummins, E.~D. Dendy, D.~B. Kothe, J.~M. Sicilian, M.~W.
  Williams, A balanced-force algorithm for continuous and sharp interfacial
  surface tension models within a volume tracking framework, J.~Comput.~Phys.
  213 (2006) 141--173.

\bibitem{Popinet_2009a}
S.~Popinet, An accurate adaptive solver for surface-tension-driven interfacial
  flows, J.~Comput.~Phys. 228~(16) (2009) 5838--5866.

\bibitem{Lopez-Herrera_2015a}
J.~L{\'o}pez-Herrera, A.~Ga{\~n}{\'a}n-Calvo, S.~Popinet, M.~Herrada,
  Electrokinetic effects in the breakup of electrified jets: A volume-of-fluid
  numerical study, Int.~J.~Multiphase Flow 71 (2015) 14--22.

\bibitem{Bell_1989a}
J.~B. Bell, P.~Colella, H.~M. Glaz, A second-order projection method for the
  incompressible {Navier-Stokes} equations, J.~Comput.~Phys. 85 (1989)
  257--283.

\bibitem{basilisk}
S.~Popinet, The basilisk code., available from http://basilisk.fr/.

\bibitem{Hooft_2018a}
J.~A. van Hooft, S.~Popinet, C.~C. van Heerwaarden, S.~J.~A. van~der Linden,
  S.~R. de~Roode, B.~J.~H. van~de Wiel, Towards adaptive grids for atmospheric
  boundary-layer simulations, Bound.-Layer Meteor. 167 (2018) 421--443.

\bibitem{Hysing_2009a}
S.-R. Hysing, S.~Turek, D.~Kuzmin, N.~Parolini, E.~Burman, S.~Ganesan,
  L.~Tobiska, Quantitative benchmark computations of two-dimensional bubble
  dynamics, Int.~J.~Numer.~Meth.~Fluids 60 (2009) 1259--1288.

\bibitem{John_2004a}
V.~John, G.~Matthies, {MooNMD}--a program package based on mapped finite
  element methods, Search Results Featured snippet from the web
  Comput.~Visual.~Sci. 6 (2004) 163--170.

\bibitem{Ganesan_2007a}
S.~Ganesan, G.~Matthies, L.~Tobiska, On spurious velocities in incompressible
  flow problems with interfaces, Comput.~Methods Appl.~Mech.~Eng. 196 (2007)
  1193--1202.

\bibitem{Menard_2007a}
T.~M\'enard, S.~Tanguy, A.~Berlemont, Coupling level set/{VOF}/ghost fluid
  methods: Validation and application to {3D} simulation of the primary
  break-up of a liquid jet, Int.~J.~Multiphase Flow 33 (2007) 510--524.

\bibitem{Jiang_2019c}
D.~Jiang, Y.~Ling, Destabilization of a planar liquid stream by a co-flowing
  turbulent gas stream, Int.~J.~Multiphase Flow 122 (2019) 103121.

\bibitem{Squire_1953a}
H.~B. Squire, Investigation of the instability of a moving liquid film,
  Br.~J.~Appl.~Phys. 4 (1953) 167.

\bibitem{Yih_1967a}
C.-S. Yih, Instability due to viscosity stratification, J.~Fluid Mech. 27
  (1967) 337--352.

\bibitem{Otto_2013a}
T.~Otto, M.~Rossi, T.~Boeck, Viscous instability of a sheared liquid-gas
  interface: Dependence on fluid properties and basic velocity profile,
  Phys.~Fluids 25 (2013) 032103.

\bibitem{Marmottant_2004a}
P.~Marmottant, E.~Villermaux, On spray formation, J.~Fluid Mech. 498 (2004)
  73--111.

\bibitem{Jarrahbashi_2016a}
D.~Jarrahbashi, W.~A. Sirignano, P.~P. Popov, F.~Hussain, Early spray
  development at high gas density: hole, ligament and bridge formations, J.
  Fluid Mech. 792 (2016) 186--231.

\bibitem{Jeong_1995a}
J.~Jeong, F.~Hussain, On the identification of a vortex, J.~Fluid Mech. 285
  (1995) 69--94.

\bibitem{Jarrahbashi_2014a}
D.~Jarrahbashi, W.~A. Sirignano, Vorticity dynamics for transient high-pressure
  liquid injection a, Phys.~Fluids 26 (2014) 73.

\bibitem{Zandian_2019a}
A.~Zandian, W.~A. Sirignano, F.~Hussain, Vorticity dynamics in a spatially
  developing liquid jet inside a co-flowing gas, J.~Fluid Mech. 877 (2019)
  429--470.

\bibitem{Sotolongo-Costa_1996a}
O.~Sotolongo-Costa, Y.~Moreno-Vega, J.~J. Lloveras-Gonz{\'a}lez, J.~C.
  Antoranz, Criticality in droplet fragmentation, Phys.~Rev.~Lett. (1996)
  42--45.

\end{thebibliography}

 \end{document}